\documentclass[a4paper,11pt]{article}
\pdfoutput=1

\usepackage[top=24truemm,bottom=34truemm,left=25truemm,right=25truemm]{geometry}

\usepackage{graphicx}
\usepackage{amsmath}
\usepackage{amssymb}
\usepackage[nosort]{cite}
\usepackage{color}
\usepackage{amsthm}
\usepackage{ulem}

\usepackage{hyperref}
\hypersetup{colorlinks,citecolor=RuriIro,linkcolor=RuriIro,urlcolor=RuriIro}
\definecolor{RuriIro}{rgb}{0.,0.28,0.60}

\numberwithin{equation}{section}
\newcommand{\bs}{\boldsymbol}

\allowdisplaybreaks

\begin{document}
\baselineskip=17.6pt

\begin{titlepage}

\begin{flushright}
KUNS-3010
\end{flushright}

\vspace*{15mm}

\begin{center}

{\Large \bfseries
Multi-Field Effects on Scalar Production in Stars}

\vspace*{8mm}

\renewcommand{\thefootnote}{\fnsymbol{footnote}}
Yasuhiro Yamamoto$^1$ and Koichi Yoshioka$^2$ 

\vspace{5mm}

{\itshape%
$^1${Physics Division, National Center for Theoretical Sciences,\\
National Taiwan University, Taipei 10617, Taiwan}\\[1mm] 
$^2${Department of Physics, Kyoto University, Kyoto 606-8502, Japan}
}%

\vspace{10mm}

\abstract{%
\noindent
This paper studies the dynamics of scalar particle production,
focusing on the presence of multiple fields and couplings in the
medium. The interplay of various fields and couplings can influence
the production rate, potentially overshadowing the effect of electrons
alone. The plasma mixing, which induces the resonance and screening of
scalar processes, is shown to be modified by the in-medium effects
depending on the type of processes and field contents. Incorporating
these in-medium effects into the analysis of stellar cooling via
scalar emission allows for the emergence of various features of
multi-field effects given in several types of scalar models.}%

\end{center}
\end{titlepage}

\renewcommand{\thefootnote}{\arabic{footnote}}
\setcounter{footnote}{0}
\setcounter{page}{1}



\section{Introduction}

In the realm of physics beyond the standard model, researchers have
investigated various avenues, including the possibility of new types
of particles, such as scalars and vectors. They have examined the
number of these particles that might exist, their potential locations
within the Universe, and the methods by which they could be
detected. A significant area of inquiry is astrophysics, where stars
serve as important testing grounds for new particle
theories~\cite{Raffelt:1996wa}. For example, the concept of using
stars to study particles has proved to be of great assistance, such as
in the case of the axion~\cite{Weinberg:1977ma,Wilczek:1977pj,Kim:1979if,Shifman:1979if,Zhitnitsky:1980tq,Dine:1981rt}, which
attempts to solve a puzzle called the strong CP
problem~\cite{Peccei:1977hh}, and as in the case of the hidden
photon~\cite{Holdom:1985ag,Okun:1982xi}, a type of particle that can
mix with the standard model photon. These
hypothetical particles could be created inside stars through processes
such as bremsstrahlung, where charged particles emit radiation as they
interact. Attempts to detect these particles in stars have contributed
to a refinement of our understanding of their properties and 
motivated experiments to detect them directly.

The CP-even scalar is another type of theoretical particle that may 
be created within stars through various processes. It often appears in
theories that go beyond the standard model, such as those involving
extended Higgs sectors or dark matter
candidates~\cite{McDonald:1993ex,Pospelov:2007mp,Fradette:2018hhl}. In
this paper, we study in detail the production of scalar particles in the
medium. As a result of their electrically neutral composition, stars
consist of multiple particles, including electrons, protons, neutrons,
and atomic nuclei. It is commonly anticipated that 
the contributions to scalar production arise from a variety of
particles with different masses, charges, and interactions with the
scalar field. In general, the influence of
heavier particles is expected to be limited and decoupled from the
main physics. For example, in bremsstrahlung and Compton scattering,
the scalar production is observed to decrease as the masses of heavier
particles increase. Consequently, it is often assumed in
discussions that only the contribution from the lightest particle
(i.e., electron) needs to be considered.

The present study examines the scalar production in the medium with a
particular focus on the roles of particles other than electrons. While
the impact of heavy particles only is indeed suppressed, it is
possible for them to give a significant effect if combined with
lighter particles (electrons). Depending on the particle energy and
couplings, this combined effect could be substantial, potentially 
dominating the production process. Therefore, it is crucial to
properly account for these characteristic effects in the medium. This
type of phenomenon, such as the plasma mixing of CP-even scalars,
leads to the screening and resonance effects of the production
rate. It has been discussed in various contexts, such as the production
of dark photons~\cite{An:2013yfc,Redondo:2013lna,Chang:2016ntp}, 
CP-even scalars~\cite{Hardy:2016kme,Dev:2020eam,Gelmini:2020xir}, and
axions in the presence of a magnetic
field~\cite{Mikheev:1998bg,Ganguly:2008kh,Caputo:2020quz}, imposing
stringent constraints on model parameters. To fully understand these
in-medium effects, it is necessary to evaluate the contributions from
multiple fields, couplings, and various processes. While the effect
of mixing and interference from multiple particles has not been
considered seriously, we demonstrate its importance by showing
significant impacts on the scalar production rate.

The subsequent sections of the paper are structured as follows: In
Section~2, we present the general expression for the scalar production
rate, including its mixing with other particles. We explore how the
scalar production process is influenced by the presence of 
various fields, couplings, and processes, especially in the context of
mixing with longitudinal photons in the medium. Furthermore, we
present a classification of these in-medium effect factors. Section~3
applies the multi-field behaviors to discuss the properties of scalar
particles inferred from the stellar cooling processes in different
scalar models. We highlight the distinct features observed in the
parameter space due to the existence of in-medium effects. Finally,
Section~4 summarizes our findings.

\bigskip

\section{Scalar production in medium}

Let us consider a scenario involving a scalar field $\phi$ (with mass
$m_\phi$) that interacts through the CP-even Yukawa couplings with matter
particles, 
\begin{align}
  {\cal L} \;=\; -y_e \phi \bar{e}e 
  -y_p \phi \bar{p}p -y_n \phi \bar{n}n \,,
  \label{eq:L}
\end{align}
where $e$, $p$, and $n$ represent the Dirac fermions corresponding to
the electron (with mass $m_e$ and charge $q_e=-e$), the proton (with charge
$q_p=e$), and the neutron, respectively. For simplicity, the masses of
the proton and the neutron are denoted by a common value $m_N$. It is
also assumed that all particles are non-relativistic and non-degenerate.

As for heavy nuclei $X$ (with atomic number $Z_X$ and mass
number $A_X$), it is assumed that their constituent nucleons can be
treated coherently, given that typical temperatures are well below the
binding energy of nuclei such as those found in stellar
environments. Consequently, throughout this paper, we consider the
mass of the nucleus to be $m_X=A_X m_N$, its charge to be $q_X=Z_X e$, and
its coupling with the scalar field $\phi$ to be 
\begin{align}
  y_X \,=\, Z_X y_p + (A_X - Z_X) y_n \,.  
  \label{eq:yX}
\end{align}

\medskip

\subsection{Production rate}

We consider the production of the bosonic field $\phi$ in a thermal
environment. The discussion of this subsection is mainly based
on~\cite{Weldon:1983jn}. For the general formalism of thermal field
theory, see e.g.,~\cite{Bellac:2011kqa,Laine:2016hma}. In a 
finite-temperature environment, as in stellar systems, the imaginary
part of the in-medium self-energy $\Pi_\phi$ is given 
by~\cite{Weldon:1983jn,Hardy:2016kme}
\begin{align}
  \text{Im}\,\Pi_\phi \,=\, -\omega \Gamma_\phi \,.
\end{align}
Here, $\omega$ is the energy of $\phi$ and the 
rate $\Gamma_\phi$ is expressed in the 
form $\Gamma_\phi=\Gamma_\text{abs}-\Gamma_\text{prod}$. For the
distribution function $f_\phi$, the rate $f_\phi\Gamma_\text{abs}$
corresponds to the decreasing of the $\phi$ number and the 
rate $(1+f_\phi)\Gamma_\text{prod}$ is the increasing one. Then, with the
principle of the detailed balance, these quantities satisfy the 
relation $\Gamma_\text{abs}=e^{\omega/T}\Gamma_\text{prod}$, where $T$
is the temperature in the medium. Equivalently, the 
relation $f_b\Gamma_\phi=\Gamma_\text{prod}$ holds for the bosonic
thermal distribution function $f_b$. In light of these properties,
according to~\cite{Hardy:2016kme}, we refer 
to $\Gamma_\text{prod}$ and $\Gamma_\text{abs}$ as the production and
absorption rates, respectively, throughout this paper. In the
subsequent section, we will evaluate the energy emission from stars by
the processes of $\phi$ production and compare it with the
observations. To this end, we will calculate the phase space integral 
of $\omega f_b \Gamma_\phi = \omega \Gamma_\text{prod}$. 

According to~\cite{Weldon:1983jn}, the production and absorption
rates of the scalar $\phi$ are expressed in terms of the corresponding
amplitudes as
\begin{align}
  \Gamma_\text{prod} &\,=\, \frac{1}{2\omega}\sum_{I,\,F}
  \int\! d\Pi_{I,F}\, |\mathcal{M}_{I\to F\phi}|^2 \,
  f_I \tilde{f}_F (2\pi)^4\delta^4(k_I-k_F-k_\phi) \,, 
  \label{eq:GammaProd}  \\
  \Gamma_\text{abs}\, &\,=\, \frac{1}{2\omega}\sum_{I,\,F}
  \int\! d\Pi_{I,F}\, |\mathcal{M}_{I\phi\to F}|^2 \,
  f_I \tilde{f}_F (2\pi)^4\delta^4(k_I+k_\phi-k_F) \,,
  \label{eq:GammaAbs}
\end{align}
where $I$ and $F$ are typically used to denote the sets of fields
participating in the processes. For the distribution function $f_X$,  
the momenta $k_X$, and their integrals $d\Pi_X$, the product/summation
are implied when multiple fields are involved in the initial and final 
states. The function $\tilde{f}_A$ is given 
by $(1\pm f_A)$ with the positive (negative) sign for the boson
(fermion). The on-shell momentum 
of $\phi$ is $k_\phi=(\omega,\bs{k})$ (hereafter represented 
as $|\bs{k}|=k$). These expressions using the 
amplitudes \eqref{eq:GammaProd} and \eqref{eq:GammaAbs} are found to 
satisfy $\Gamma_\text{abs}=e^{\omega/T}\Gamma_\text{prod}$, and 
the imaginary part of boson self-energy (the discontinuity) can be
expressed with the amplitudes~\cite{Weldon:1983jn} 
via $\text{Im}\,\Pi_\phi=-\omega(\Gamma_\text{abs}-\Gamma_\text{prod})$.

We then turn our attention to the case in which the scalar $\phi$ has 
the mixing self-energy with other fields. Our focus is on the $\phi$
production, including the processes by which it is generated via the
mixing with photons in the medium. This type of production phenomenon
has been explored for various bosons: the dark
photon~\cite{An:2013yfc,Redondo:2013lna}, the CP-even
scalar~\cite{Hardy:2016kme}, and the axion in the presence of a
background magnetic 
field~\cite{Mikheev:1998bg,Ganguly:2008kh,Caputo:2020quz}. These
studies have imposed notable constraints on the properties of
these bosons. If $\phi$ mixes with the longitudinal 
photon $\gamma_L^{}$ (the plasmon), the resulting sum of all contributions 
through the mixing leads to the self-energy
\begin{align}
  \Pi_\phi \,=\, 
  \Pi_{\phi\phi} - \frac{\Pi_{\phi L}\Pi_{L\phi}}{\Pi_{LL}-K^2} \,.
  \label{eq:Piphi}
\end{align}
See Appendix~A for the derivation. 
Here $\Pi_{AB}$ denotes the sum of all one-particle irreducible loop
contributions to the $AB$ two-point functions, and $K$ is the
(generally off-shell) four-momentum of external particles. 
For the imaginary part of self-energy, the previously referenced
expression in terms of amplitudes is now applicable 
to $\text{Im}\,\Pi_{\phi\phi}$. A parallel expression also exists for 
$\text{Im}\,\Pi_{LL}$, with its production term written by
\begin{align}
  \Gamma_\text{$L,\,$prod} &\,=\, \frac{1}{2\omega}\sum_{I,\,F}
  \int\! d\Pi_{I,F}\, |\mathcal{M}_{I\to FL}|^2 \,
  f_I \tilde{f}_F (2\pi)^4\delta^4(k_I-k_F-k_L) \,,
\end{align}
using the corresponding amplitudes for the plasmon with the 
momentum $k_L$. The absorption term is not explicitly given here, as
it is not required for the following analyses. The imaginary part of
the self-energy of $\phi$ in the presence of mixing \eqref{eq:Piphi}
can be found by using the amplitudes introduced above.

We assume that the fields in the theory couple both to $\phi$ and the
photon (if this is not the case, the corresponding charge is formally
taken as zero), and that $\phi$ has the mixing self-energy 
with $\gamma_L$ via these couplings. By taking the imaginary part of
the right-hand side of \eqref{eq:Piphi} and making a straightforward
mathematical modification using the amplitudes, we find that the
production part in the presence of mixing can be rewritten as follows:
\begin{align}
  \Gamma_\text{prod} \,=\, \frac{1}{2\omega}\sum_{I,\,F} 
  \int\! d\Pi_{I,F}\, |\mathcal{M}_{I\to F\phi}^\text{eff}|^2 
  f_I \tilde{f}_F (2\pi)^4\delta^4(k_I-k_F-k_\phi) \,,
\end{align}
where the effective $\phi$ production amplitude in the medium is
defined by taking into account the mixing with plasmons,
\begin{align}
  \mathcal{M}_{I\to F\phi}^\text{eff} \,=\,
  \mathcal{M}_{I\to F\phi} \,-\, 
  \frac{\Pi_{L\phi}}{\Pi_{LL}-K^2}\, \mathcal{M}_{I\to FL} \,.
  \label{eq:Meff}
\end{align}
The plasmon momentum is taken to be $k_L=k_\phi$. According to this
expression, the $\phi$ production rate in the medium can be obtained by
replacing $\mathcal{M}_{I\to F\phi}$ with 
$\mathcal{M}_{I\to F\phi}^\text{eff}$ in the expression of the vacuum
amplitude, i.e., the amplitude in the absence of the plasmon mixing. 

Ref.~\cite{Hardy:2016kme} evaluates the scalar production rate from
the imaginary part of $\Gamma_\phi$ (and multiplied by $f_b$). They 
explicitly calculate the leading real parts from thermal field theory
and evaluate the whole part, including the imaginary ones, by using
the proportionalities among the leading results. That presents a
promising approach for assessing the effects from a single
field/coupling. In the case of multiple fields, which will be
discussed later, there are some difficulties to deal with where
the forms of self-energies would be different from those obtained by
assuming a simple proportionality relation in the single
field/coupling case. To provide a prescription for studying such a
multi-field case, we introduce the expression in terms of
corresponding amplitudes, following~\cite{Weldon:1983jn}. Instead of
summing up all production processes, we will estimate the structure of
amplitudes based on proportional relations, similar to 
what \cite{Hardy:2016kme} does, and calculate the production rate. The
above form rewritten with amplitudes is extending the method of
Ref.~\cite{Hardy:2016kme} to make it easier to handle the case of
multiple fields.

\medskip

\subsection{In-medium effects}

The effective production amplitude \eqref{eq:Meff} in the medium can
be expressed in the following form
\begin{align}
  |\mathcal{M}_{I\to F\phi}^\text{eff}|^2 \,=\, 
  z_\phi\, c_\phi^{IF} |\mathcal{M}_{I\to F\phi}|^2 \,,
  \label{eq:zcM}
\end{align}
\begin{align}
  z_\phi &\,=\, \bigg|\frac{K^2}{K^2-\Pi_{LL}}\bigg|^2 \,,  \\
  c_\phi^{IF} &\,=\, 
  \Big|1+\frac{1}{K^2\,\mathcal{M}_{I\to F\phi}}
  \big( \Pi_{L\phi}\mathcal{M}_{I\to FL} - 
     \Pi_{LL}\mathcal{M}_{I\to F\phi} \big) \Big|^2 \,.
  \label{eq:cphi}
\end{align}
The amplitude is divided into two components: the vacuum amplitude
in the absence of mixing and two types of in-medium effect factors,
$z_\phi$ and $c_\phi$. The superscripts of the multi-field 
effect $c_\phi$ generally denote the fields appearing in the process. The
factor $z_\phi$ commonly appears for all processes.

Firstly, $z_\phi$ represents the common medium effect in the presence
of mixing with plasmons. By parametrizing $\Pi_{LL}$ with the
renormalizing factor as 
$\text{Re}\,\Pi_{LL}=(K^2/\omega^2)\omega_p^2$ and 
$\text{Im}\,\Pi_{LL}=-(K^2/\omega) \Gamma_L$~\cite{Redondo:2013lna}, 
$z_\phi$ becomes
\begin{align}
  z_\phi \,=\,
  \frac{\omega^4}{(\omega^2-\omega_p^2)^2+\omega^2\Gamma_L^2} \,,
  \label{eq:zphi}
\end{align}
where $\omega_p$ denotes the plasma frequency in the medium, 
and $\Gamma_L$ corresponds to the damping rate of the plasmon. As can
be seen from the expression of \eqref{eq:zphi}, $z_\phi$ emerges as a
universal factor associated with any process, regardless of the
specific properties of $\phi$ and its production processes. With
regard to the energy dependence, $z_\phi$ is almost $1$ in the
high-energy regime $\omega>\omega_p$ that indicates the suppression of
the plasmon effect. In the vicinity of the resonance 
at $\omega\sim\omega_p$, the localized amplification of $\phi$
production arises, potentially yielding the dominant
contribution~\cite{Hardy:2016kme}. Conversely, in the low-energy
regime where $\omega<\omega_p$, the suppression occurs with respect 
to $\omega$, resulting in the screening of $\phi$
processes~\cite{Gelmini:2020xir}. Defining $\omega_c$ by 
$\omega_c\Gamma_L(\omega_c)=\omega_p^2$, there exist two distinct
regions of screening: $z_\phi\sim(\omega/\omega_p)^4$ for $\omega>\omega_c$
and $z_\phi\sim (\omega/\Gamma_L)^2$ for $\omega<\omega_c$. 

The factor $c_\phi$ captures the influence of multiple fields/multiple
couplings in the medium.  As will be seen later, $c_\phi^{IF}=1$ for
any $\phi$ process in the case of a single field/single coupling. This
is because the second term in \eqref{eq:cphi} cancels out and 
there is no medium effect attributed to $c_\phi$. In general, the value 
of $c_\phi$ varies depending on the $\phi$ processes. The second term
in \eqref{eq:cphi} inversely correlates with the energy $\omega$ and
increases in the lower-energy region compared to $\omega_p$,
potentially becoming a significant factor. This behavior contrasts
with that of $z_\phi$, which induces the screening of processes
at low energy. In conclusion, according to \eqref{eq:zcM}, the
production rate of the scalar $\phi$ in the medium can be expressed as
the product of (the ubiquitous factor $z_\phi$) $\times$ (the variable
factor $c_\phi$ for each process) $\times$ (the production rate in the
vacuum).

\medskip

\subsubsection{Single field : only $z_\phi$}

We assume particles in the medium to be non-relativistic and
non-degenerate. In this case, their couplings to $\phi$ via Yukawa
coupling $y$ and to $A^0$ via electromagnetic coupling $q$ exhibit the
same form at the leading order. Furthermore, we
observe that $\Pi_{\phi L}$ satisfies $\Pi_{\phi L} = 
\varepsilon_{L\,\mu}\,\Pi_{\phi A}^\mu 
=(-\sqrt{K^2}/k)\Pi_{\phi A}^0$ through the Ward-Takahashi
identity. Here $\varepsilon_L$ represents the polarization vector of
the longitudinal mode, $\varepsilon_L(K)=\tfrac{1}{\sqrt{K^2}}
(k,\omega\boldsymbol{k}/k)$. Consequently, the ratio of coupling
strength between a certain $\phi$ process and its corresponding $\gamma_L$
process is generally given by $y/q_L$, where we 
define $q_L=(-\sqrt{K^2}/k)\,q$.

As the simplest case, we consider a charged field with a single
Yukawa coupling $y$. This corresponds to the situation where only the
effect of electrons is considered in the medium. For the production 
process $I\to F+\phi\;(\gamma_L)$ driven by the coupling $y$ ($q$),
the amplitudes satisfy
\begin{align}
  \mathcal{M}_{I\to F\phi} \,:\, \mathcal{M}_{I\to FL} 
  \;=\; y \,:\, q_L \,.
  \label{eq:yqL}  
\end{align}
Moreover, given the presence of only one type of field and coupling,
we observe that
\begin{align}
  \Pi_{L\phi} \,:\, \Pi_{LL}  \;=\; y \,:\, q_L 
  \label{eq:PiLLLphi}
\end{align}
as well. Consequently, as seen in \eqref{eq:cphi}, the factor
$c_\phi^{IF}$ becomes equal to 1 in this case, regardless of the 
detail of processes such as $I$ and $F$. This is why $c_\phi$ is
referred to as the multi-field effect. The cancellation of the second
term in \eqref{eq:cphi} indicates the existence of a field basis in which 
(non-relativistic) electrons and $\phi$ do not interact. By taking an
appropriate linear combination of $\phi$ and the plasmon depending on the
coupling ratio $y/q_L$, one can define a new basis where the scalar
$\phi'$ has no Yukawa coupling. Then the direct $\Pi_{\phi'\phi'}$
becomes suppressed and the scalar is only generated through the mass
mixing with the plasmon. Note that such a coupling-dependent field
redefinition is generally not useful in the case of multiple fields
with non-uniform coupling ratios. This suggests a possibility of
important effects beyond only $z_\phi$, as discussed in the subsequent
section.

In the end, the production rate of $\phi$ for the case of a single
field/coupling is obtained by the product of the medium effect $z_\phi$
and the production rate in the absence of mixing
$\Gamma_\text{prod}^{(\text{no mix})}$. According to the relation 
\eqref{eq:yqL}, $\Gamma_\text{prod}^{(\text{no mix})}$ 
correlates to the production rate
of plasmon $\Gamma_\text{$L,\,$prod}$ which 
is determined by $\Gamma_L$ in the single field/coupling case, namely, 
$\Gamma_\text{prod}^{(\text{no mix})} 
=(y/q_L)^2\Gamma_\text{$L,\,$prod}
=(y/q)^2(k^2/\omega^2)\,f_b\Gamma_L$. Then, the $\phi$ production
rate takes the form
\begin{align}
  \Gamma_\text{prod} \,=\,\frac{y^2}{q^2}\,
  \frac{f_b k^2\omega^2 \Gamma_L}{(\omega^2-\omega_p^2)^2
+\omega^2\Gamma_L^2} \,,
\end{align}
which is the well-known expression~\cite{Hardy:2016kme} for the $\phi$
production in the medium. It should be noted that throughout this
expression, the energy and momentum are with respect to $\phi$. 

It is necessary to address several points. Firstly, if we 
consider $\Gamma_L$ to be the actual damping rate of plasmon, it 
should cover the contributions from all processes. However, not all of
these processes necessarily follow \eqref{eq:yqL}. Secondly, when
calculating the damping rate of plasmon, particularly in stellar
environments, there exists a contribution from the 
electron-nucleon bremsstrahlung process. Even if only one charged field
has the Yukawa coupling to $\phi$, the relation \eqref{eq:PiLLLphi} is
modified by the presence of multiple charged fields, e.g., the
electron and the nucleons. In any case, the cancellation in $c_\phi$ 
discussed above does not occur, and it is likely that the in-medium
effects extend beyond just $z_\phi$.

\medskip

\subsubsection{Multiple fields : $z_\phi$ and $c_\phi$}

As explained above, $z_\phi$ is the common factor that is independent
of the production processes. In contrast, $c_\phi$ determines the
specific characteristics of the production rate for each process. The
evaluation of $c_\phi$ requires the evaluation of $\Pi_{L\phi}$, which
can be challenging in the general case beyond the simple scenario
with a single field/coupling. In the following discussion, we
proceed under the assumption $\Pi_{L\phi}(K)=\Pi_{\phi L}(K)$. In this
situation, for any $I$ and $F$, the phase of $\mathcal{M}_{I\to F\phi}$ is 
equal to that of $\mathcal{M}_{I\to FL}$ where the external $\phi$
line is replaced with $\gamma_L$. This enables us to 
express $\Pi_{L\phi}$ from the relations between the amplitudes. 

Let us consider the situation where two types of charged fields are
present in the medium: the electron $e$ with the charge $q_e$ and a
single type of atomic nucleus $X$ with the charge $q_X$ and
mass $m_X$. The number densities of these particles in the
medium are denoted by $n_e$ and $n_X$, respectively. In the context of
stellar environments, which will be the focus of the subsequent section, 
we study the bremsstrahlung and Compton processes for the production
of $\phi$. These processes can be classified as follows: 
(i) the emission from electrons: $e$-$e$ bremsstrahlung, $e\gamma$
Compton, etc., 
(ii) the emission from nuclei: $X$-$X$ bremsstrahlung, $X\gamma$
Compton, etc., and 
(iii) the emission from both electrons and nuclei $X$: $e$-$X$
bremsstrahlung, etc. 
Each of these processes exhibits distinct in-medium effects,
reflecting the influence of multiple fields and couplings. While the 
medium effects of process (iii) can be derived by combining those of
(i) and (ii), we here present the case of (iii) independently, since it
serves as a representative production process. According 
to \eqref{eq:cphi}, we find the in-medium effect 
factors $c_\phi^e$, $c_\phi^X$, and $c_\phi^{eX}$ for the processes
(i), (ii), and (iii) as follows:
\begin{align}
  c_\phi^e &\,=\, \Big|\,1+\frac{1}{\omega^2}\,
  \frac{q_ey_X - q_Xy_e}{y_e} \Big[ 
  \frac{q_Xn_X}{m_X}  -i \frac{q_X}{m_X^2} P_X 
  +\frac{i}{m_X} \Big(\frac{q_e}{m_e}
  -\frac{q_X}{m_X}\Big)P_{eX} \Big]\,\Big|^2  \,,
  \label{eq:cphie}  \\
  c_\phi^X &\,=\, \Big|\,1-\frac{1}{\omega^2}\,
  \frac{q_ey_X - q_Xy_e}{y_X}
  \Big[ \frac{q_e n_e}{m_e} 
  -i \frac{q_e}{m_e^2} P_e - \frac{i}{m_e}
  \Big(\frac{q_e}{m_e}-\frac{q_X}{m_X}\Big)P_{eX} \Big]\,\Big|^2  \,,
  \label{eq:cphiX}  \\
  c_\phi^{eX} &=\, \Big|\,
  1+\frac{1}{\omega^2}\frac{q_ey_X-q_Xy_e}{y_e m_X-y_X m_e}
  \Big[ q_en_e+q_Xn_X - i\frac{q_e}{m_e}P_e -i\frac{q_X}{m_X} P_X 
  \Big]\,\Big|^2  \,.
  \label{eq:cphieX}
\end{align}
The functions $P$'s are defined as
\begin{align}
  P_i &\,=\, \frac{q_i^4 n_i^2}{60\pi^{3/2} m_i^{1/2}}
\,\frac{T^{1/2}J(\omega)}{f_\phi \omega^2}  +\frac{q_i^2 n_i k}{6\pi} \,,
  \label{eq:Pi}  \\
  P_{ij} &\,=\, 
\frac{q_i^2q_j^2n_in_j m_i^{1/2}m_j^{1/2}}{6\sqrt{2}\pi^{3/2}(m_i+m_j)^{1/2}}
\,\frac{I(\omega)}{f_\phi T^{1/2}\omega^2} \,,
  \label{eq:Pij}
\end{align}
where $T$ is the temperature in the medium. $P_i$ means the
contribution from the $i$ field, while $P_{ij}$ represents the
contributions related to the $i$-$j$ bremsstrahlung. These functions
correspond to the imaginary parts of the in-medium self-energies. The
dimensionless functions $I(\omega)$ and $J(\omega)$ are given by
\begin{align}
  I(\omega) &\,=\, 
  \int^\infty_{\tfrac{m_\phi}{T}} du \int^\infty_0 \! dv 
  \int^1_{-1}\!dz \, \sqrt{uv}\,e^{-u} 
  \,\frac{\delta(u-v-\omega/T)}{u+v-2\sqrt{uv}\,z}  \,,   \\
  J(\omega) &\,=\,
  \int^\infty_{\tfrac{m_\phi}{T}} du \int^\infty_0\! dv 
  \int^1_{-1}\! dz \, \sqrt{uv}\,e^{-u} \,\delta(u-v-\omega/T)
\frac{(u+v)^2+12uvz^2}{\big[(u+v)^2-4uvz^2\big]^2}  \nonumber \\
  &\hspace*{35mm} 
  \times \Big[\,
  3\big[(u+v)^2-4uvz^2\big]\Big(1-\frac{m_\phi^2}{\omega^2}\Big) 
+\frac{4\omega^2}{T^2}+\frac{m_\phi^2}{T^2} \,\Big] \,.
  \label{eq:fncJ}
\end{align}
The comments on the in-medium factors
\eqref{eq:cphie}--\eqref{eq:cphieX} are in order: 

(i) $c_\phi^A$ does not include the contribution from $P_A$
($A=i,\,j,\>ij$). This is analogous to the case of a single
field/coupling where self-contributions cancel out in the second term
of $c_\phi$ \eqref{eq:cphi}. For example, if a process $A$ yields the
most substantial contribution, the corresponding $P_A$ may naturally
give a significant effect. However this large $P_A$ does not affect
the in-medium factor $c_\phi^A$ for the process $A$ itself, while it
has a large impact on processes other than $A$.

(ii) The real part of $c_\phi^{eX}$ \eqref{eq:cphieX} is determined by
the sum of $q_en_e$ and $q_Xn_X$. In the environment consisting of two
particles $e$ and $X$, this contribution is cancelled out to zero due
to the condition of electrical neutrality. The cancellation occurs
because, under the neutral condition in the medium, the self-energies
of non-relativistic particles can be written in the forms proportional
to the dipole-like combinations of Yukawa and electromagnetic
couplings, which is analogous to the case of bremsstrahlung amplitudes.

(iii) All multi-field effects are proportional to the combination of
couplings in the form of $q_ey_X-q_Xy_e$. In other words, the effects
are manifest only through a conspiracy of two types of fields, and
hence called the multi-field effect. A charged field without 
Yukawa coupling to $\phi$ can contribute to $c_\phi$ (via other charged
fields), while a neutral field with Yukawa coupling also contributes
to $c_\phi$. Another important point is that multi-field effects are
absent if the electromagnetic charge $q_i$ is universally proportional
to $y_i$. For example, consider the presence of extra gauge bosons
such as the dark photon with kinetic mixing to the standard model
photon. In this case, the extra boson couplings $y_i$ satisfy the 
relation $y_i=(\text{const})q_i$, regardless of the type of charged
fields. Analogous evaluation of the in-medium effect factors reveals
that multi-field effects are absent for the extra boson
production~\cite{Hardy:2016kme} (see also \cite{Li:2023vpv} for
deviations from the universality). The vanishing of multi-field effects
can also be understood from the field redefinition discussed in the
previous section. In the presence of multiple fields, if the 
ratios $y_i/q_i$ have a common value (that is, $q_iy_j=q_jy_i$), all
multi-field effects disappear through a certain redefinition with a single
rotation angle which is determined by this ratio.

The plasmon damping rate and the plasma frequency are found to be
\begin{align}
  \Gamma_L &\,=\, 
  \frac{1}{\omega} \Big(\frac{q_e}{m_e}-\frac{q_X}{m_X}\Big)^2P_{eX} 
  +\frac{q_e^2}{\omega m_e^2}P_e  +\frac{q_X^2}{\omega m_X^2}P_X \,,
  \label{eq:gammaL}  \\
  \omega_p^2 &\,=\, \frac{q_e^2n_e}{m_e}+\frac{q_X^2n_X}{m_X} \,.
  \label{eq:omegap}
\end{align}
Taking in mind the electrically neutral condition in the medium, 
the dominant contribution to $\omega_p$ is given by the electrons, while
the contributions from other nuclei are always suppressed by the
hierarchical mass ratio $m_e/m_N$. Nevertheless, it should be noted
that even if the contribution of heavy particles is suppressed, it may
still give a dominant physical effect in cases that the leading-order
cancellation occurs, as described above.

\medskip

\subsection{Production rates with in-medium effects}
\label{sec:zphicphi}

We analyze the property of $\phi$ production rates, taking into
account both of the in-medium effect factors $z_\phi$ and $c_\phi$. We
focus on the case with two species of particles in the 
medium: the electron ($e$) and the hydrogen nucleus (H). As illustrative
examples, we study three types of processes, $e$-$e$ bremsstrahlung, H-H
bremsstrahlung, and $e$-H bremsstrahlung.

Figure~\ref{fig:zphicphi} (the left panel) shows the typical in-medium
factors $z_\phi$ and $c_\phi$. The medium environment, such as the 
temperature and the number density, is chosen to be $T=1$~keV 
and $n_e=10^{26}$~cm$^{-3}$ which are similar to the solar interior. The
Yukawa couplings are set to be $y_e=10^{-9}$ and 
$y_H/m_H=2y_e/9m_e$. The latter follows the relation 
in~\cite{Shifman:1979eb}, consistent with the Higgs portal model where
the scalar $\phi$ couples to 
the Higgs boson through the renormalizable interaction and gives 
a simple description of new physics and wide phenomenological and
experimental motivations~(for example,~\cite{Patt:2006fw,Beacham:2019nyx}).
The overall magnitude of Yukawa 
couplings does not change the qualitative behavior of the lines. The
scalar mass $m_\phi$ is assumed to be sufficiently small, rendering
its effect negligible in the figure. For larger 
values of $m_\phi$ compared to the plasma frequency, all in-medium
effects such as the resonance and the screening disappear. The following
provides a brief elucidation of $z_\phi$ and the three types of
multiple-field factors $c_\phi$: 
\begin{figure}[t]
\centering
\includegraphics[width=7.7cm]{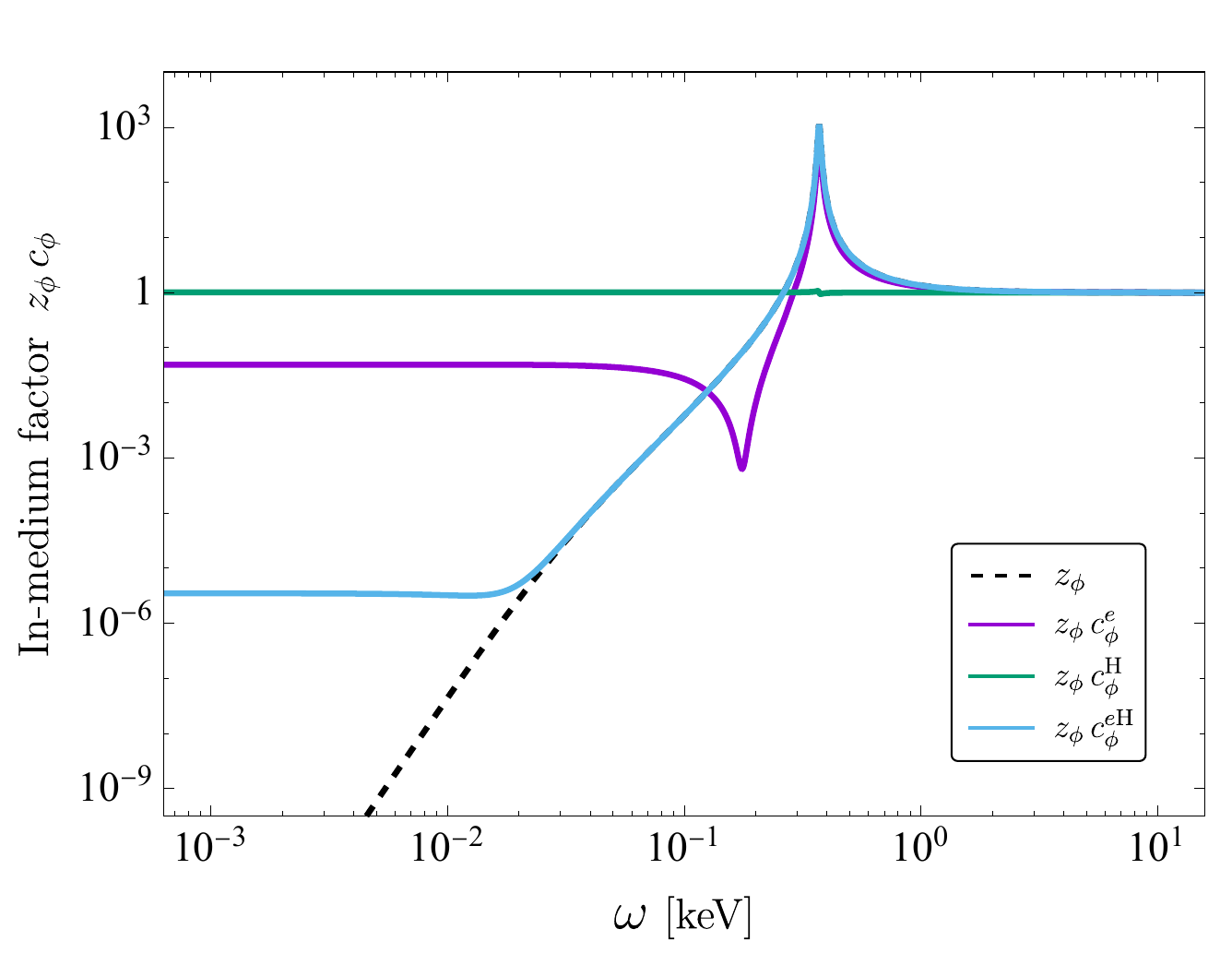}
~
\includegraphics[width=7.8cm]{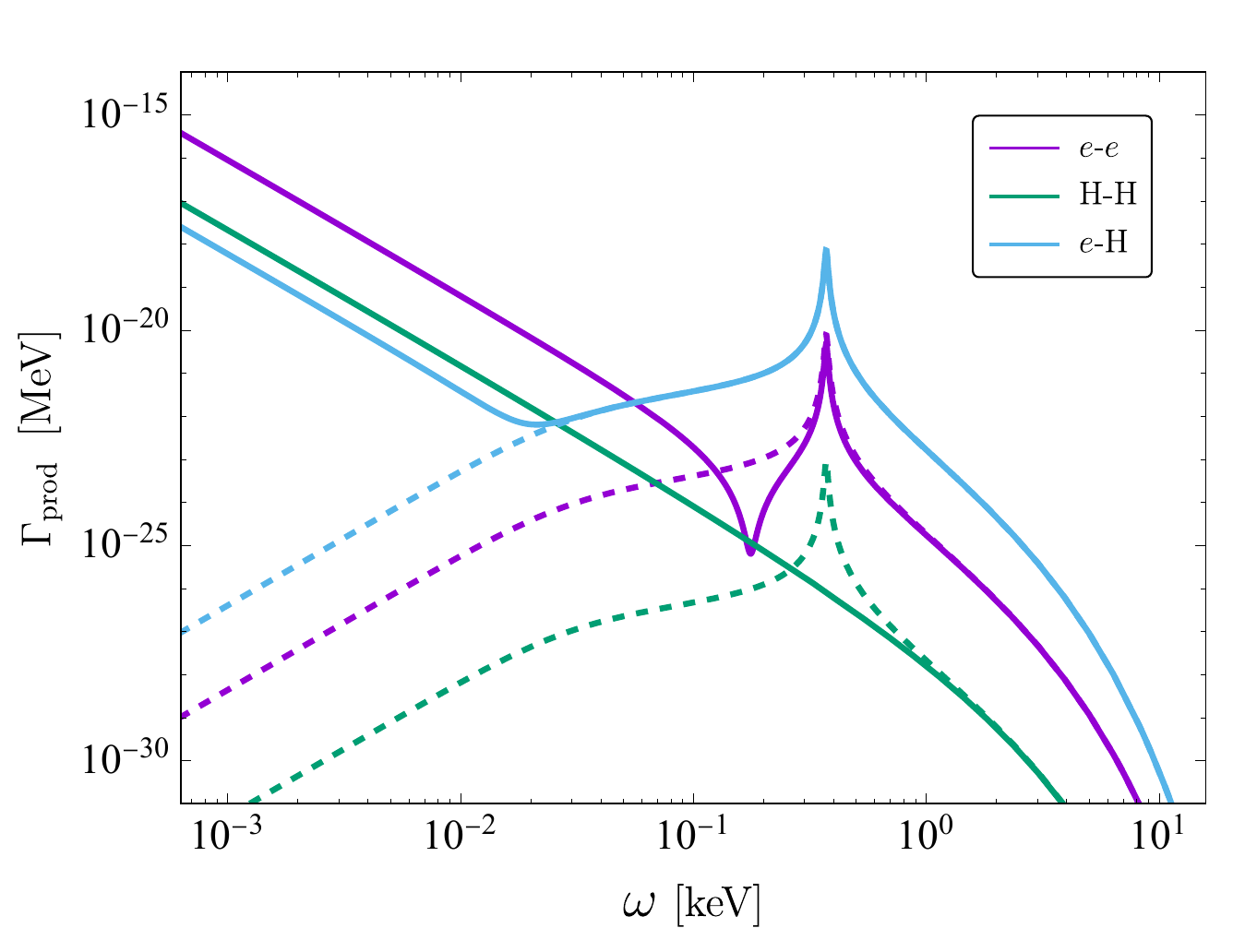}
\caption{
(Left) The in-medium effect factors $z_\phi$
and $c_\phi$. The purple, green, and blue lines correspond to the
factors of $ee$ type, HH type, and $e$H type, respectively. The black
dashed line represents $z_\phi$ alone. 
(Right) The production rates with the in-medium effect factors. Three
types of bremsstrahlung are depicted: $e$-$e$ (purple), H-H (green),
and $e$-H (blue). The temperature and the electron number density are set
to the solar-like values, specifically $T=1$~keV 
and $n_e=10^{26}$~cm$^{-3}$. The Yukawa couplings are chosen to 
be $y_e=10^{-9}$ and $y_H/m_H=2y_e/9m_e$ as in the Higgs portal
model. The qualitative behavior of the lines is independent of the
overall magnitude of Yukawa couplings.}
\label{fig:zphicphi}
\bigskip
\end{figure}

\begin{itemize}
\item 
The dashed black line, which partially overlaps with the blue line,
illustrates $z_\phi$. The deviations from this dashed line indicate the
effects of multiple fields and couplings, which are represented 
by $c_\phi$. As discussed in \eqref{eq:zphi}, the behavior of $z_\phi$ is
characterized by the negligible effect ($z_\phi\simeq1$) at high
energy, the resonance ($z_\phi\gg1$) near the plasma 
frequency $\omega_p$, and the screening ($z_\phi\ll1$) at low
energy. The behavior of $z_\phi$ remains almost unchanged even in the 
presence of multiple fields and couplings. As previously discussed, the
plasma frequency is dominantly influenced by the lightest particle,
the electron, regardless of the Yukawa couplings with $\phi$.

\item 
For the $ee$ type (purple line), the impact of multi-field effects is 
negligible in the high-energy regime ($c_\phi^e\sim1$). In the low-energy
regime, however, the screening effect of $z_\phi$ is partially
countered by $c_\phi^e$, leading to a deviation from the dashed black
line. This weakening of the screening is attributable to the
contributions from the real part of $c_\phi^e$ \eqref{eq:cphie} 
for $\omega>\omega_c$ and to the contributions from the imaginary part
of the $e$-H type process (the term $P_{eX}$) for $\omega<\omega_c$.

Furthermore, in the $ee$ type, a dip-shaped feature is observed near
the scale where the real part of $c_\phi^e$ \eqref{eq:cphie} becomes
zero. The location of this dip is approximately found to 
be $\omega\sim (y_Nm_e/y_em_N)^{1/2}\omega_p$. 

\item 
For the HH type (green line), the influence of $z_\phi$ is entirely
offset by $c_\phi$. This is due to the real part of $c_\phi$ for
$\omega_c\lesssim\omega\lesssim\omega_p$ and the imaginary
part for $\omega\lesssim\omega_c$. Consequently, 
$z_\phi c_\phi\simeq 1$ holds over the entire energy range, which
eliminates the medium effects including the characteristic behaviors 
of resonance and screening. Thus, the $\phi$ production
amplitude is identical to that in the vacuum. This is mainly due to
the fact that the plasma frequency is determined by the electron
contribution and with no significant effect from other nuclei. The
conclusion does not depend on the magnitude of hydrogen Yukawa
coupling $y_\text{H}$ as long as $y_\text{H}\gtrsim y_e$. 

\item 
For the $e$H type (blue line), in the high-energy regime 
above $\omega_c$, the influence of multiple fields is absent and the
medium effect is solely determined by $z_\phi$. Consequently, the
resonance and screening of production rates appear. Below $\omega_c$,
the cancellation of screening takes place due to the multi-field
effect $c_\phi$, resulting in a constant medium factor. A weaker
cancellation compared to the $ee$ and HH types can be 
explained by the self-cancellation within $c_\phi$, as previously
discussed. Among the three bremsstrahlung processes under 
consideration, the $e$H type has the largest amplitude in the vacuum
due to the hierarchical mass ratio $m_e/m_\text{H}$ and the Pauli
blocking, which in turn makes it the weakest in the medium through
the multi-field effect. The cancellation of screening in the $e$H
type is attributed to the contribution of $ee$ bremsstrahlung. As a
result, the blue line shows the suppression of approximately 
$(T/m_e)^2\sim10^{-5}$ compared to the cancellation observed in 
the $ee$ type (the terminal value of the purple line). 

\end{itemize}
We further comment that there is little variation in $z_\phi$ and
$c_\phi$ as a result of differences in medium properties. The only
somewhat visible modification involves the shift in the plasma
frequency (the location of resonance) and the temperature-dependent
cancellation in the $e$-H bremsstrahlung. The other dependencies, such
as those on Yukawa couplings and the species of nuclei, can be derived
from the general expressions~\eqref{eq:cphie}--\eqref{eq:cphieX}.

\bigskip

The right panel of Figure~\ref{fig:zphicphi} shows the $\phi$ 
production rates including the in-medium factors. Three types
of production processes, the $e$-$e$, H-H, and $e$-H bremsstrahlung, are
represented by the purple, green, and blue lines, respectively. Based
on the kinetic theory, i.e., without the medium effects, the
production rate is known to simply behave as $\omega^{-3}$. In
addition to this monotonic behavior, the inclusion of the universal
factor $z_\phi$ only is shown by the dashed line for each color,
exhibiting the resonance and screening 
phenomena~\cite{Hardy:2016kme,Gelmini:2020xir}. The additional
inclusion of the multi-field factor $c_\phi$ yields the solid line
for each color. The behavior of the production rate in the medium is
thus determined by the product of 
$z_\phi c_\phi \times \text{(the
monotonic behavior of kinetic theory)}$ and reflects the
characteristics of $z_\phi$ and $c_\phi$ discussed above. 

For example, the H-H bremsstrahlung process shows the identical
behavior in the medium and in the vacuum, which matches to the kinetic
theory result and is unaffected by the resonance or screening
effect. Furthermore, one can see the low-energy screening effect is
cancelled for any process, thereby restoring the energy dependence in
the vacuum. The low-energy values of production rates vary for each
process due to the differences in the extent of screening
cancellation. For example, the $e$-H bremsstrahlung, which has the
largest rate in the vacuum, experiences the weakest cancellation, and
then the $e$-$e$ bremsstrahlung becomes the dominant production rate
at lower energy. Consequently, in terms of the production number, the
$e$-$e$ bremsstrahlung may exceed the resonance. (If the absorption of
$\phi$ in the medium is taken into account, the production rate saturates 
at the energy where the rate is roughly equal to typical medium size.)
~It is also mentioned that as the resonance is universally governed
by $z_\phi$, the most significant bremsstrahlung process in the
vacuum, i.e., the $e$-H one, also dominantly contributes to the resonance.

\bigskip

\section{Stellar cooling}

The existence of extra particles interacting with the standard
model fields, such as a new scalar and an extra photon, results in the
emission of these particles which serve as additional sources of
energy loss. This phenomenon is observed in astrophysical
objects, which provides various constraints on the properties of
these extra particles~\cite{Raffelt:1996wa}. In the following,
we will discuss the constraints on the CP-even scalar $\phi$ obtained
from the stellar cooling argument, with particular focus on the Sun 
and horizontal-branch (HB) stars. So far, various single-field
analyses have been performed in previous literature, see, for
example~\cite{Hardy:2016kme,Dev:2020eam,Bottaro:2023gep,Yamamoto:2023zlu}. 
In the medium of stars, the total electric charge should be neutral in the
presence of electrons and nucleons as well as atomic nuclei formed by
their accumulation. When the CP-even scalar $\phi$ couples 
to these multiple fields with different charges, masses, and
couplings, the stellar cooling via the $\phi$ emission can be evaluated
by using the results in the previous section.

\medskip

\subsection{Cooling processes and emission rates}

In this work, we suppose that the Sun is composed of electrons,
hydrogen, and helium. The HB stars are similarly assumed to be
composed of electrons and helium for simplicity. We present several
specific expressions of the production rates for electrons and helium
(corresponding to the HB stars), while the production rates involving
multiple species of nuclei (like the Sun) are given in Appendix B.

When the scalar $\phi$ generally has the Yukawa couplings to
electrons and nucleons as in the Lagrangian \eqref{eq:L},
the following $\phi$ production processes can be effective at the
energy scale in stars (now HB stars):
(i) $e$-$e$ bremsstrahlung : $e+e\to e+e+\phi$, 
~(ii) He-He bremsstrahlung :
$\text{He}+\text{He}\to\text{He}+\text{He}+\phi$, 
~(iii) $e$-He bremsstrahlung : $e+\text{He}\to e+\text{He}+\phi$, 
~(iv) $e$ Compton-like~: $e+\gamma\to e+\phi$, and
(v) He Compton-like : $\text{He}+\gamma\to \text{He}+\phi$.
The processes (i) and (ii) receive the suppression of $T/\text{mass}$
compared to (iii) due to the cancellation resulting from the existence 
of identical particles in the final states. We nevertheless include
them since they can still be dominant processes in some cases, as
demonstrated by the multi-field effects discussed in the previous
section. The above bremsstrahlung processes are all
photon-mediated. In the process (ii), there is also the pion-mediated
one, which is not included because of its suppression 
by $(T/m_N)^2$ compared to the photon-mediated CP-even scalar
production. Moreover, the emission of $\phi$ from the mediator 
photon can also be neglected as being small. For example, when the
coupling between $\phi$ and $\gamma$ is induced by standard-model
fermion loops, the ratio of production rates is 
approximately ($\gamma$ emit)/($e$ emit) $\sim (T/m_e)(\alpha/4\pi)^2 \ll 1$.

Assuming non-relativistic electrons and nucleons, the production rates
for these processes including the full in-medium effects are found to be
\begin{align}
  \Gamma_\text{prod}^{ee} &\,=\,  z_\phi c_\phi^e
  \frac{e^4 y_e^2 n_e^2 T^{1/2}k^2J(\omega)}{60\pi^{3/2} m_e^{5/2}\omega^5}\,,
  \\
  \Gamma_\text{prod}^\text{HeHe} &\,=\,  z_\phi c_\phi^\text{He}
  \frac{2e^4 y_N^2 n_\text{He}^2 
T^{1/2}k^2J(\omega)}{15\pi^{3/2} m_N^{5/2}\omega^5} \,,  \\
  \Gamma_\text{prod}^{e\text{He}} &\,=\,  z_\phi c_\phi^{e\text{He}}
 \frac{\sqrt{2}e^4 n_e n_\text{He}^{} m_e^{1/2}}{3\pi^{3/2}}
  \Big(\frac{y_e}{m_e} -\frac{y_N^{}}{m_N}\Big)^2
 \frac{k^2 I(\omega)}{T^{1/2}\omega^5}  \,,
  \label{eq:eHebrems}   \\
  \Gamma_\text{prod}^{e\gamma} &\,=\, 
  z_\phi c_\phi^e \frac{e^2 y_e^2 n_e f_\phi k^3}{6\pi m_e^2 \omega^3} \,,
  \\
  \Gamma_\text{prod}^{\text{He}\gamma} &\,=\, 
  z_\phi c_\phi^\text{He} \frac{2e^2 y_N^2 n_\text{He}
f_\phi k^3}{3\pi m_N^2 \omega^3}  \,.
\end{align}
The medium effect $z_\phi$ is given by \eqref{eq:zphi}, and
$c_\phi$'s and the plasma frequency are obtained by substituting the
masses and coupling constants in 
\eqref{eq:cphie}--\eqref{eq:cphieX} and \eqref{eq:omegap} by those of 
electron and helium 
$q_e=-e$, $q_\text{He}=2e$, $m_\text{He}=4m_N$, and $y_\text{He}^{}=
2y_p+2y_n\equiv 4y_N^{}$. In HB stars with the composition ratio
assumed above, the number density satisfies $n_\text{He}=n_e/2$
according to the condition of electrical neutrality, and hence the
real part of $c_\phi^{e\text{He}}$ cancels out. In the above
production rates, the contributions of heavy nuclei enter via the
Yukawa couplings $y_{p,n}$ and are suppressed by the nucleon mass,
with the suppression factor being $y_{p,n}/m_N$. In the case of a
scalar coupled to the standard model through the mixing with the Higgs
boson, referred to as the Higgs portal scalar in the following, the
Yukawa coupling is generally proportional to the corresponding
mass. Consequently, the contribution of heavy nuclei is not
negligible since $y_X/m_X\sim y_{p,n}/m_N \sim y_e/m_e$. Furthermore,
the effects of heavy nuclei are also observed through the multi-field
effects $c_\phi$. These effects can give characteristic influences on
the scalar production, as previously discussed.

\begin{figure}[t]
\centering
\includegraphics[width=7.7cm]{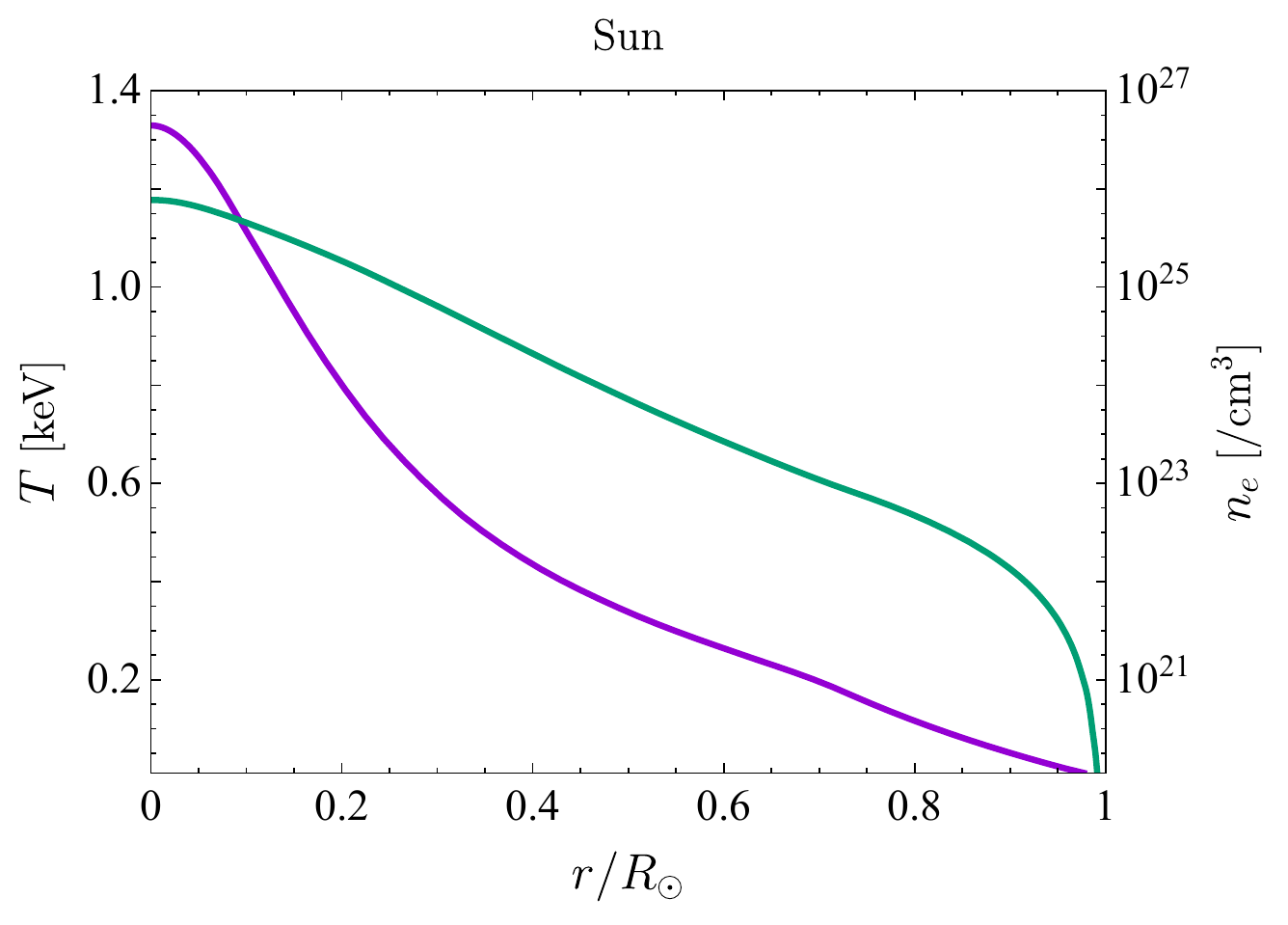}
\includegraphics[width=7.7cm]{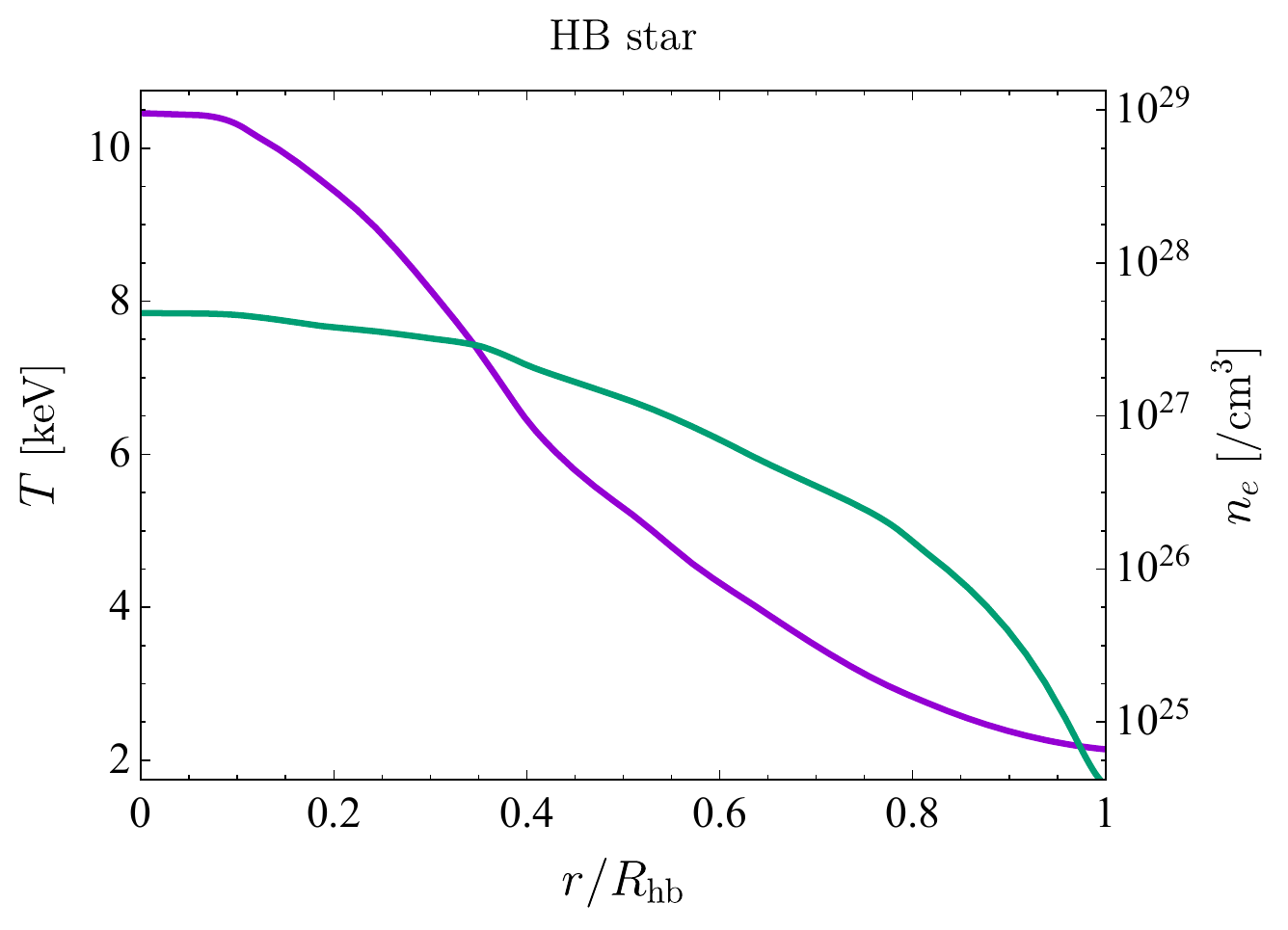}
\caption{Typical profiles of the temperature (purple line) and the 
electron number density (green line) inside stars: (left) the 
Sun~\cite{Vinyoles:2016djt,Li:2023vpv} and (right) HB stars (data taken
from~\cite{Raffelt:1996wa}). The solar radius is 
$R_\odot=7\times 10^{10}$ cm. For HB stars, we here choose the radius as
$R_\text{hb}=5.4\times 10^9$ cm within which the helium processes are
important.}
\label{fig:profile}
\bigskip
\end{figure}
The energy emission rate per unit time (luminosity) from
stars is evaluated by 
\begin{align}
  L^i \,=\, \int\! d^3\bs{r} \!\int\! \frac{d^3\bs{k}}{(2\pi)^3}\,
  \omega\,\Gamma_\text{prod}^i
  \,=\, \frac{1}{2\pi^2} \!\int_0^R\!\! dr\,r^2 \! \int\! d^3\bs{k}\,
  \omega\,\Gamma_\text{prod}^i(T(r),n(r))
  \label{eq:Q}
\end{align}
for each process. We assume the internal structure of stars to be
spherically symmetric. Here, $r$ and $R$ represent the distance from
the center and the radius of a star, respectively. In the above
expression, the opacity factor denoting the scalar absorption in the
medium has been dropped, as this has no significant effect in the
parameter regions of interest. The profiles of the temperature $T(r)$
and the electron number density $n_e(r)$ for the Sun and HB stars are
shown in Figure~\ref{fig:profile}. As a simplified model for
evaluating the luminosity, the Sun is 
assumed to have the mass composition of hydrogen and helium with the  
ratio of $3:1$, and HB stars are considered to consist solely of 
helium. In HB stars, depending on the degree of evolution,
carbon and oxygen may also be present in the core, but we neglect
their effects for simplicity of calculation. While there is also the 
region of hydrogen burning outside the helium-dominated portion
(referred to in Figure~\ref{fig:profile} as within the 
radius $R_\text{hb}=5.4\times 10^9$~cm), the effect there on the stellar
energy loss through $\phi$ emission is negligible due to the low
number density and temperature.

\begin{figure}[t]
\centering
\includegraphics[width=6.5cm]{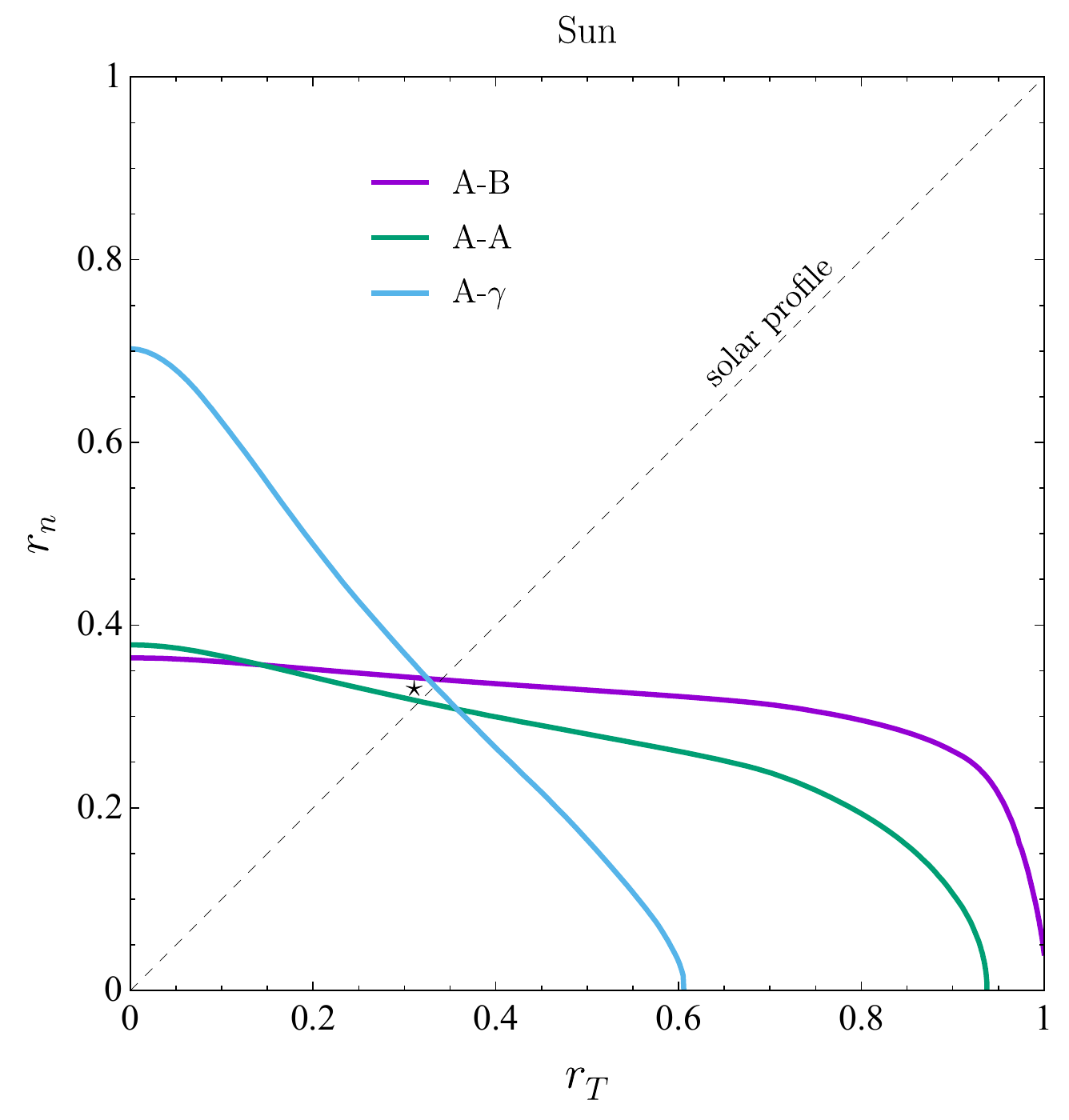}
\hspace*{10mm}
\includegraphics[width=6.5cm]{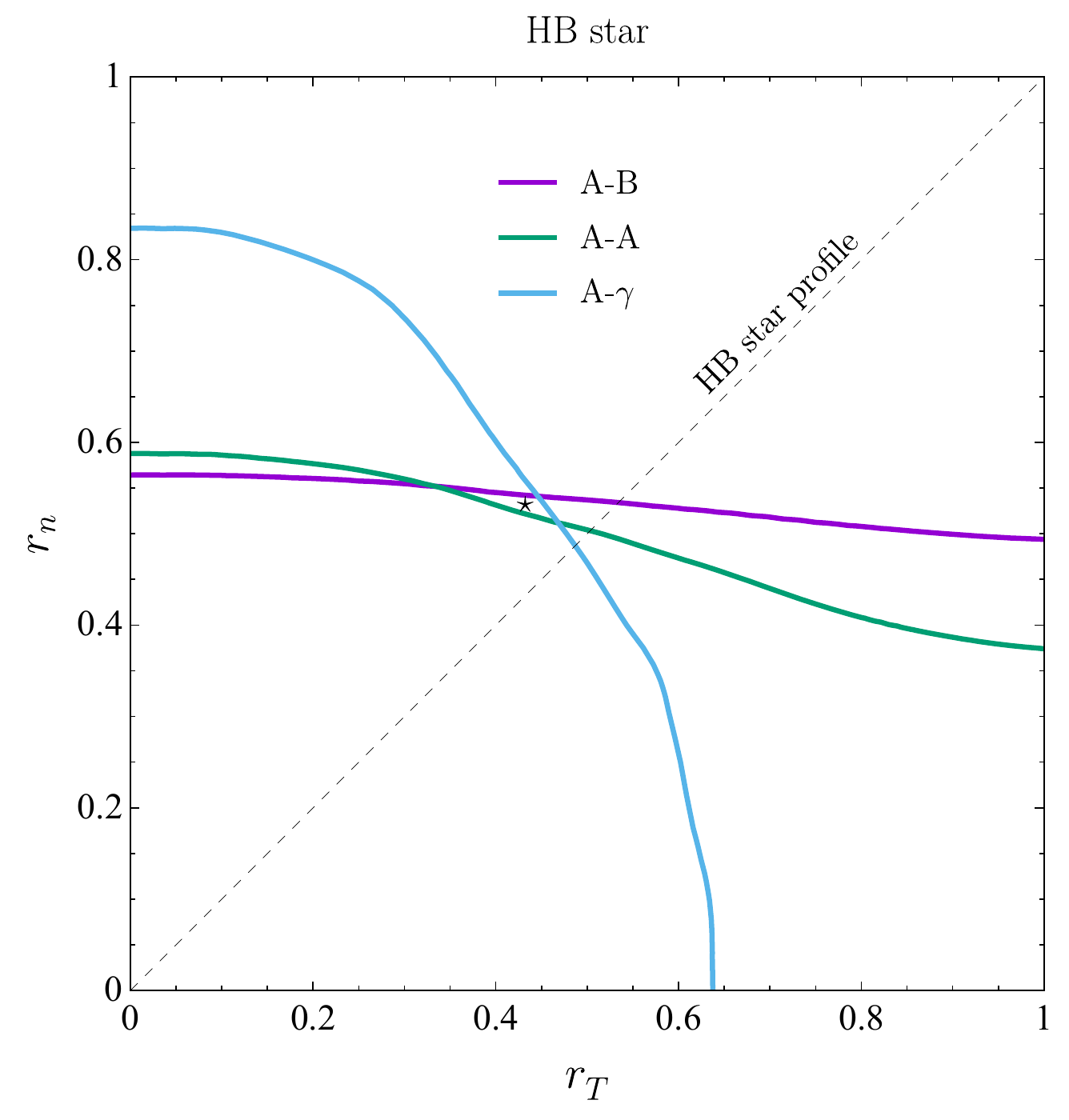}
\caption{Uniform temperature and density capturing 
the luminosity using the profiles of 
Figure~\ref{fig:profile}. The horizontal and vertical axes correspond
to the temperature $T(r_T^{}R)$ and the density $n_i(r_n^{}R)$,
respectively, where $R$ is the radius of each star. The purple, green,
and blue lines represent the comparisons with the A-B type 
bremsstrahlung, the A-A type bremsstrahlung, and the A-$\gamma$ Compton-like
processes, respectively. (left) The Sun and (right) HB stars. The black
stars correspond to $T_*=0.56$~keV 
and $n_{e*}=4.6\times10^{24}$~cm$^{-3}$ on the left, and $T_*=6$~keV 
and $n_{e*}=1.1\times10^{27}$~cm$^{-3}$ on the right.}
\label{fig:Qeq}
\bigskip
\end{figure}

The luminosity \eqref{eq:Q} is generally evaluated by using the
profiles. Here we adopt a convenient prescription for quantitative
evaluation~\cite{Yamamoto:2023zlu} where certain uniform temperature
$T_*$ and density $n_*$ can mimic the evaluation using the
profiles. There are three types of energy 
emission processes: the A-B bremsstrahlung, the A-A bremsstrahlung, and
the A-$\gamma$ Compton-like (A, B = electron, nucleus). For these
processes, we show in Figure~\ref{fig:Qeq} the uniform temperature and
density at which the luminosities match to \eqref{eq:Q} using the
profiles given in Figure~\ref{fig:profile}.
The uniform temperature $T_*$ and density $n_{e*}$ are parametrized
by $r_T$ and $r_n$ using the profile functions such 
that $T_*=T(r_T^{}R)$ and $n_*=n(r_n^{}R)$ where $R$ is the radius of
each star. The left panel corresponds to the Sun, while the right
panel shows the results for HB stars. It is found in both cases that
the lines for three types of processes nearly coincide at a single 
point. That indicates that the evaluation using the profile can
be reproduced by the uniform temperature and density. In the following 
numerical analysis, we adopt this prescription and use the uniform 
temperature and density marked with the black stars in
Figure~\ref{fig:Qeq}, which specifically correspond to
\begin{alignat}{2}
\text{~~~Sun :} & \quad T_* \,=\, 0.56~\text{keV} \,,\qquad 
&& n_{e*} \,=\, 4.6\times 10^{24}~\text{cm}^{-3} \,,  \\
\text{HB star :} & \quad T_* \,=\, 6~\text{keV} \,, 
&& n_{e*} \,=\, 1.1\times 10^{27}~\text{cm}^{-3} \,.
\end{alignat}

\medskip

\subsection{Leptonic/hadronic couplings}

Consider the case that the scalar $\phi$ has Yukawa couplings 
both to the electrons and nucleons. In the stellar medium, both
the electrons and nuclei are present, capable of producing $\phi$ and
emitting the stellar energy. The Yukawa coupling with nucleons is
considered isoscalar, which means it is common both to the 
protons and neutrons. Consequently, the Yukawa coupling
with the nucleus $X$ is given by $y_X=A_Xy_N^{}$, with $A_X$ being the
mass number of $X$. The present model parameters then consist of two
Yukawa couplings, one for leptonic and one for hadronic 
interactions, $y_e$ and $y_N$, respectively. By comparing the
production of a hypothetical scalar $\phi$ with the observational
data, the limits on these couplings can be discussed in the context of
stellar cooling. In the analysis of this subsection, we set the scalar
mass to a typical value of $m_\phi=10^{-2}$~keV.

\begin{figure}[t]
\centering
\includegraphics[width=6.8cm]{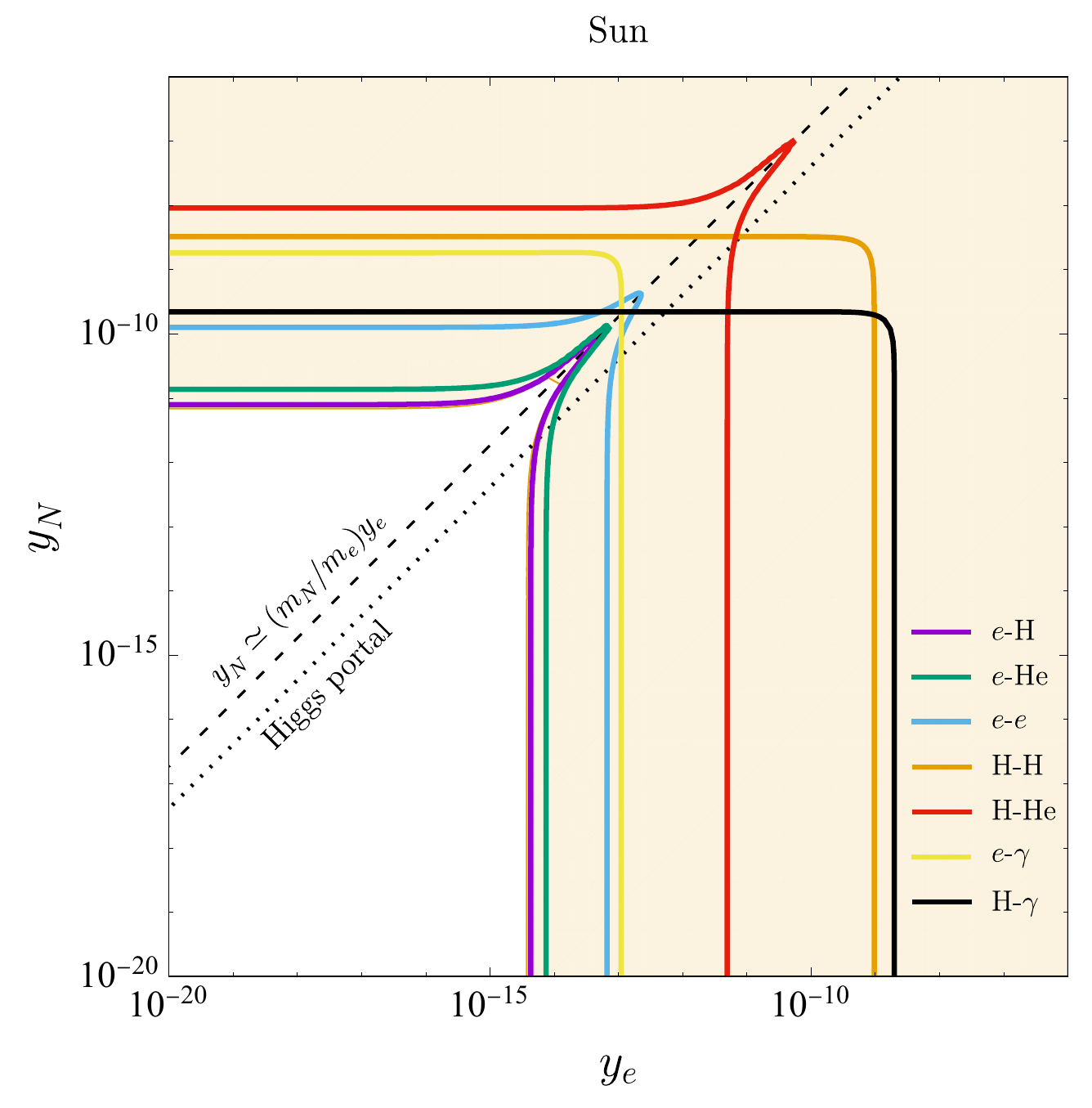}
~~~~
\includegraphics[width=6.8cm]{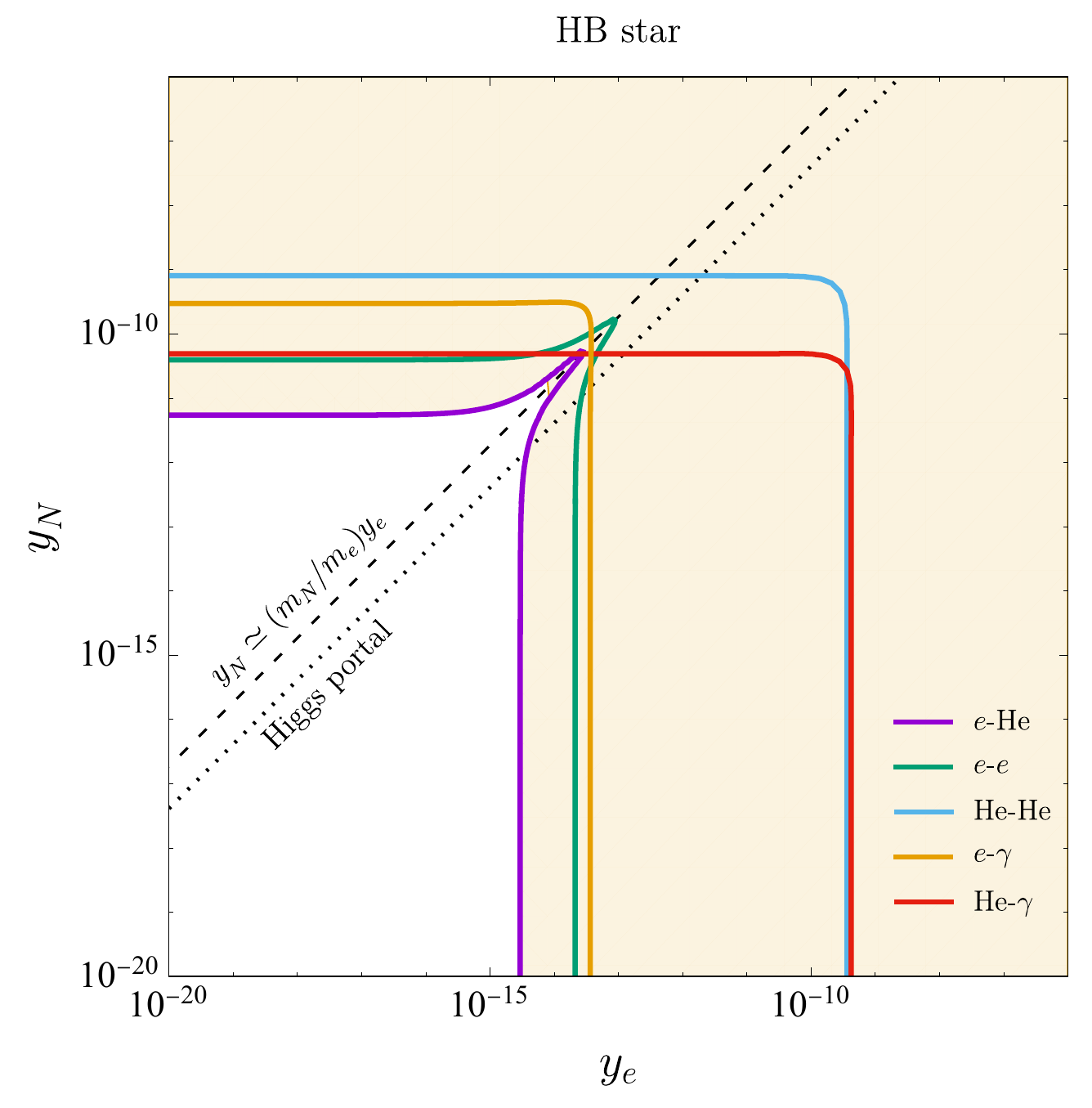}
\caption{The bounds on scalar leptonic/hadronic Yukawa couplings from
the stellar cooling. (left) The Sun and (right) HB stars. The orange
shaded area represents the excluded region, which comes from the
requirement that the scalar luminosity should be less than $0.03$
($5$) of the solar luminosity in the left (right)
panel. $m_\phi=10^{-2}$~keV is assumed. The black dotted lines
represent the case of the Higgs portal scalar. See the text for details
on the dent feature along the black dashed lines.}
\label{fig:yeyN}
\bigskip
\end{figure}

The bounds on the leptonic/hadronic scalar couplings resulting from
stellar cooling are illustrated in Figure~\ref{fig:yeyN}.
The shaded regions represent the excluded parameter regions where the
scalar luminosity exceeds 0.03 of the solar luminosity leading to
the excessive cooling in the solar case, and exceeds 5 times the solar
luminosity in the HB star case. The former value is used
here as a typical upper limit determined by the maximum amount of
extra cooling, when analyzing solar 
cooling for the axion~\cite{Vinyoles:2015aba}, and the level of extra
cooling would remain relatively consistent regardless of particle
type. The latter is chosen conservatively based on the typical
luminosity of HB stars and the experimental accuracy of their
lifetime~\cite{Raffelt:1996wa,Ayala:2014pea}. 
If a different upper limit needs to be set, the results obtained
here can be roughly modified accordingly.  
Each colored line in Figure~\ref{fig:yeyN} means the boundary that
gives the constraint solely from the cooling with that process.

As a specific model, the Higgs portal scalar is shown by the black
dotted lines in the figure~\cite{Hardy:2016kme,Yamamoto:2023zlu}. There, 
the bremsstrahlung scalar emission from the electron has dominant effects,
imposing the most stringent constraints. The bounds 
for hadrophilic (only scalar-nucleon coupling $y_N$) and leptophilic
(only scalar-electron coupling $y_e$) scenarios correspond
to the upper left and lower right regions of the figure,
respectively. In these regions, we find the upper bounds, specifically, 
$y_N<8.0\times 10^{-12}$ and $y_e<4.3\times 10^{-15}$ for the Sun, and 
$y_N<5.4\times 10^{-12}$ and $y_e<2.9\times 10^{-15}$ for the HB 
stars. These bounds are in approximate agreement with those
derived in the hadrophilic and leptophilic limits of the $e$-nucleus
bremsstrahlung process in HB stars~\cite{Hardy:2016kme}.

The existence of multiple fields in the medium, the electron and
nucleon, gives rise to several characteristics. Firstly, even
electron-like processes such as the $e$-$e$ bremsstrahlung 
and $e\gamma$ Compton impose the bounds on the nucleon
coupling. Note that, without including multi-field 
effects, $L^{ee}$ and $L^{e\gamma}$ provide no bounds 
on $y_N$. Conversely, even nucleon-like processes such as 
the $X$-$X$ bremsstrahlung and $X\gamma$ Compton impose the bounds on
the electron coupling through the multi-field effect
$c_\phi^X$. Secondly, one can see in the figure that some processes
exhibit local weakening of constraints. For example, in the case 
of $e$-He bremsstrahlung, the production rate in the
absence of mixing is observed to decrease 
around $y_e/m_e \simeq y_N/m_N$, resulting in the formation of
a sharp dent in the ($y_e,\,y_N$) plane. In the case
of $e$-$e$ bremsstrahlung, the dominant contribution comes from the
plasmon mixing in the medium which arises from the $e$-He bremsstrahlung.
However it cancels out the vacuum contribution in the absence of
mixing near the resonance, at which the $\phi$ emission is most
effective. This situation can be expressed by 
using the condition of electrical neutrality as 
\begin{align}
  \text{Re}\;c_\phi^e|_{\omega\simeq\omega_p} \,\simeq\, 0 
  \quad\to\quad  y_N/y_e +1/2\,\simeq\, m_N/m_e \,,
\end{align}
which corresponds approximately to the dashed black lines in
Figure~\ref{fig:yeyN}. It should be noted that there is no long
extending slit seen in these directions in the parameter space. The
presence of multi-field effects results in the slits being limited to
some extent and only gives dents. It can also be observed that the
dent regions are covered by other processes in addition to the 
multi-field effects.

\medskip

\subsection{Hadrophilic scalar in stars}

As another application to examine the contribution of heavy particles
to scalar production in the medium, the next attention is directed to
the Yukawa coupling between the scalar and nucleons. In this
subsection, we consider the hadrophilic scenario with negligible
$y_e$, which means that the model parameters are the scalar-proton
coupling $y_p$ and the scalar-neutron coupling $y_n$. The scalar-quark
couplings are high-energy model dependent and generally involve the
isospin breaking. When matched to the effective theory of nucleons,
they generally induce isovector components, and hence the Yukawa
couplings of protons and neutrons can be different even for the signs
(see, for example, \cite{GrillidiCortona:2015jxo} for the axion
case). We assume that the scalar coupling of a nucleus $X$ is given by
the general form \eqref{eq:yX} and  the scalar mass is set to 
$m_\phi=10^{-2}$~keV, as in the previous subsection.

We first note that a characteristic process in the presence of
multiple nuclei is the scalar production via the $X$-$X'$
bremsstrahlung ($X\neq X'$). For the previously assumed  
composition, this process is absent in the HB stars. Firstly, the
production rate of this $X$-$X'$ bremsstrahlung in the vacuum without
the plasmon mixing is typically large. We find a rough 
estimation, $\Gamma_\text{prod}^{XX'}/\Gamma_\text{prod}^{eX}
\sim Z_X (m_X/m_e)^{1/2}\gg 1$, under the assumption of electrical
neutrality. Even in the non-hadrophilic 
case, $\Gamma_\text{prod}^{XX'}$ can be dominant if $y_X$ is
sufficiently large, $y_X/y_e\gtrsim (m_X/m_e)^{3/4}$. Secondly, the
non-relativistic amplitude of $X$-$X'$ bremsstrahlung in the vacuum 
potentially decreases due to the cancellation. 
When the masses of the proton and neutron are approximately taken to be
equal, the production amplitude of $X$-$X'$ bremsstrahlung
satisfies (see, for example, Appendix \eqref{eq:XXpBrems})
\begin{align}
  \mathcal{M}_{XX'\to XX'\phi} \,\propto\, 
  \Big(\frac{A_X}{Z_X}-\frac{A_{X'}}{Z_{X'}}\Big)(y_p-y_n) \,.
\end{align}
Thus the vacuum amplitude vanishes when (a) the Yukawa couplings are
universal for the nucleons ($y_p=y_n$), and (b) the ratios $A/Z$ are
universal for the nuclei (typically $A/Z=2$). The case (a) was assumed
in the previous subsection, which resulted in a considerable weakening
of the constraint from $L^\text{HHe}$ at the Sun, as shown in
Figure~\ref{fig:yeyN}. It is noted that if the vacuum amplitude
becomes zero, non-trivial contributions via plasmon mixing 
through $c_\phi$ generally exist in the medium, and some constraints
can be obtained. In this subsection, we have non-universal nucleon 
couplings, but the contribution of $X$-$X'$ bremsstrahlung can also be
vanished if the condition (b) is met. For example, HB stars are
assumed to be helium-dominated and satisfy the 
condition. This conclusion remains unchanged even when the presence of
carbon and oxygen is considered, since their $A/Z$ ratios are still
2. However, the condition does not apply to the medium containing 
hydrogen ($A/Z=1$) like in the Sun and in the material containing certain
isotopes. In the end, the production rate of $X$-$X'$
bremsstrahlung in the vacuum is generally larger than those of other
processes and may be an important factor without the above cancellations.

\begin{figure}[t]
\centering
\includegraphics[width=6.8cm]{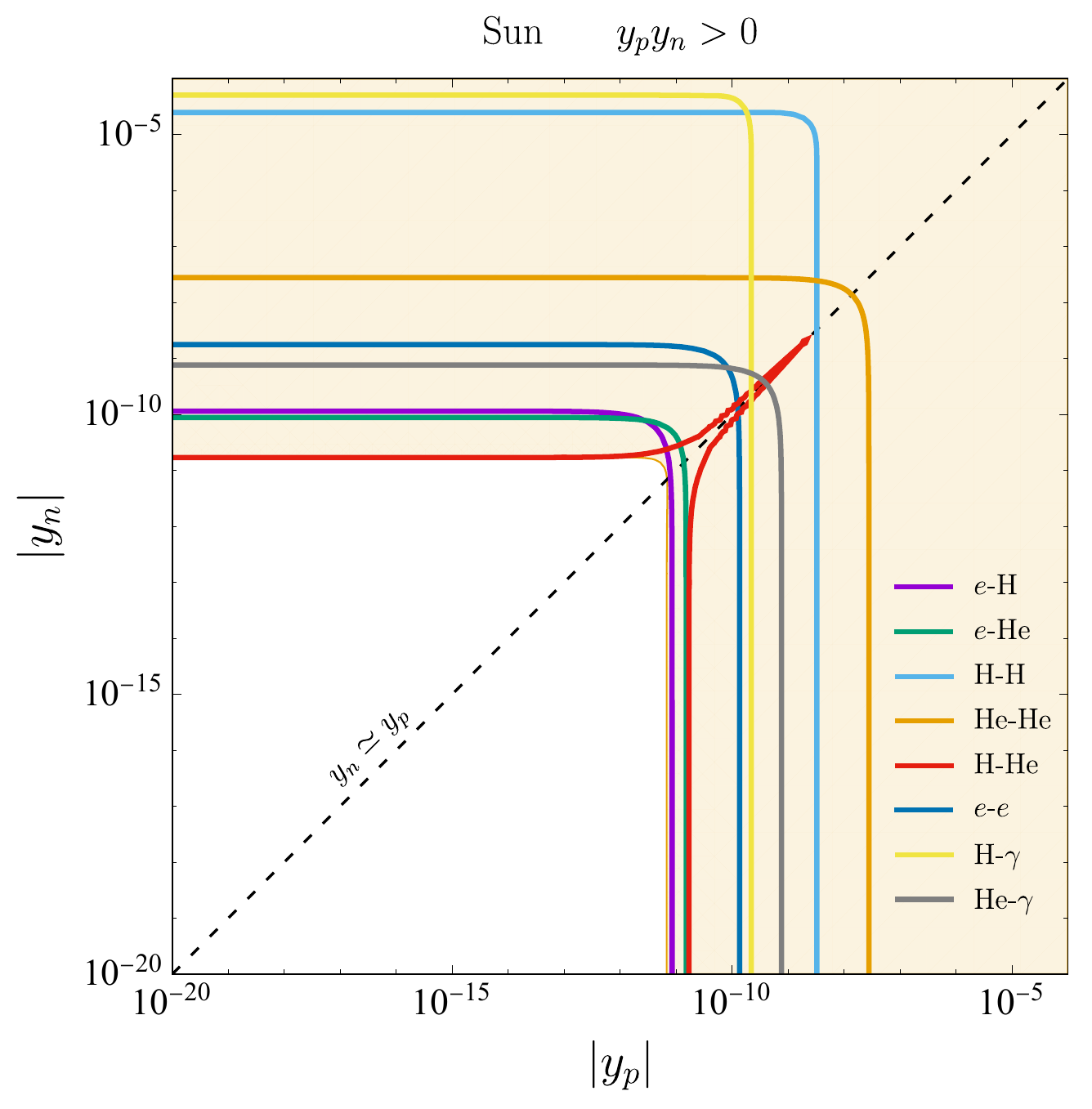}
~~~~
\includegraphics[width=6.8cm]{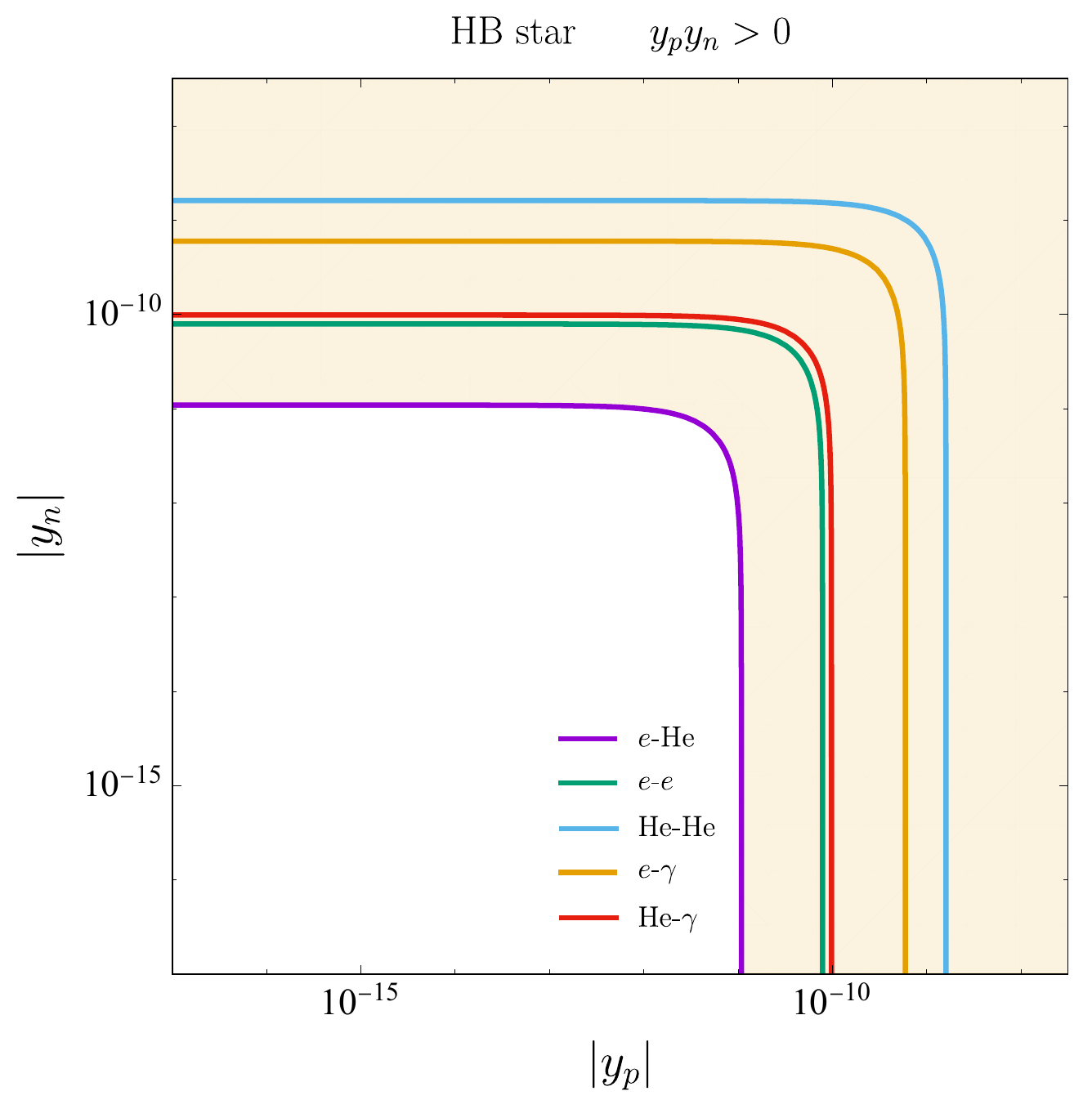}
\bigskip

\includegraphics[width=6.8cm]{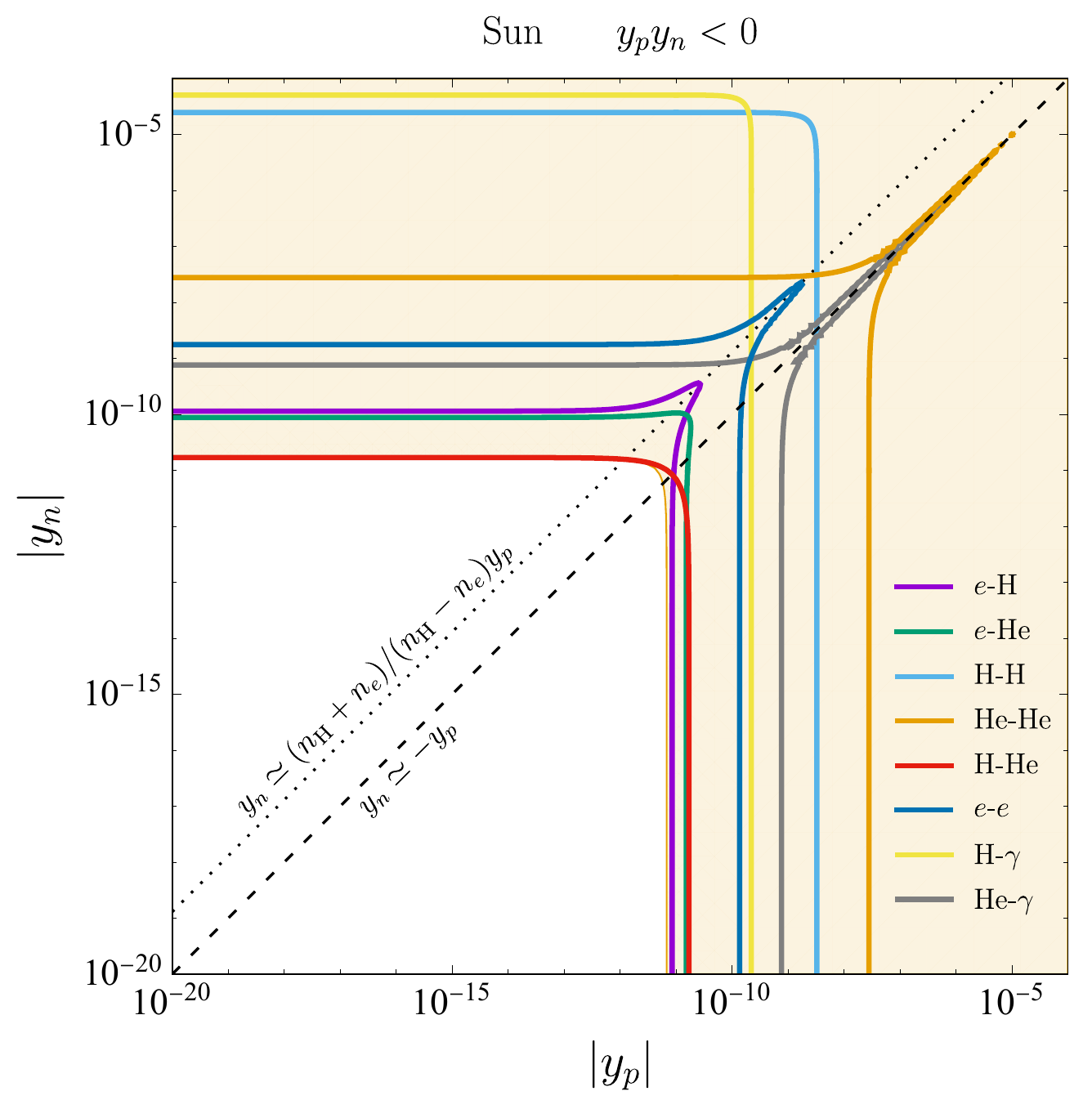}
~~~~
\includegraphics[width=6.8cm]{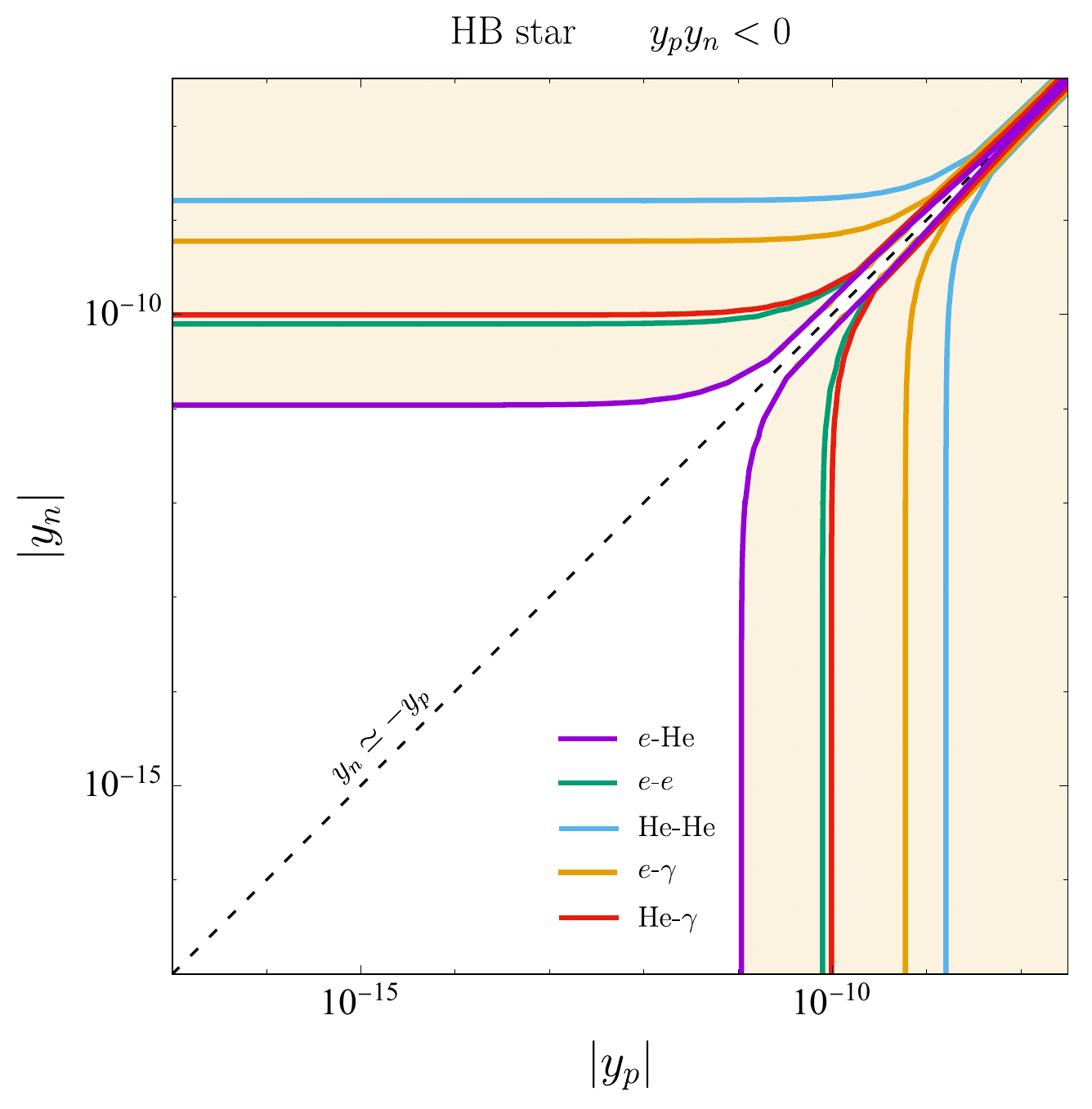}
\caption{The bounds on the hadrophilic scalar couplings from the stellar
cooling: (top) $y_p$ and $y_n$ are of the same sign and (bottom)
they are of opposite signs. (left) The Sun and (right) HB
stars. The orange shaded area represents the excluded regions,
indicating that the scalar luminosity should be less than $0.03$ ($5$)
times the solar luminosity for the Sun (HB
stars). $m_\phi=10^{-2}$~keV is assumed. As for the dashed and dotted
lines, see the text for details.}
\label{fig:ypyn}
\bigskip
\end{figure}

Figure~\ref{fig:ypyn} shows the constraints on the hadronic Yukawa
parameters from the stellar cooling argument. One can see, even in the
case of hadrophilic scalar, the parameter limits about nucleon
couplings could be potentially derived from the contributions of
electron-like processes such as the $e$-$e$ bremsstrahlung via the
multi-field medium effects. The limit from the $e\gamma$ Compton has
not been shown for the Sun as it is relatively weak, at least weaker
than that from $L^{ee}$.

We begin by discussing the qualitative behaviors in the case of HB
stars (the right two panels in Figure~\ref{fig:ypyn}). Since only
helium is effective and its Yukawa coupling appears as the 
combination $y_\text{He}^{}=2(y_p+y_n)$, the constraints on 
the $(y_p,\,y_n)$ plane become symmetric 
under $y_p\leftrightarrow y_n$. In addition, a sharp slit is observed
in the vicinity of $y_p\simeq -y_n$, indicating that there is no
parameter limit in such a hadrophilic scalar model (the stellaphobic
scalar). This behavior of the stellaphobic scalar is not altered even
if there exist different or multiple nuclei in the stellar
medium. This is because, in the absence of hydrogen or isotopes as
effective elements for the scalar production, the couplings only
appear as the combination $y_p+y_n$, regardless of the inclusion of
multi-field medium effects.

In the case of the Sun, as hydrogen is present in addition to helium,
the situation is rather different and complicated (the left two panels in
Figure~\ref{fig:ypyn}). One can see several sharp dents and slits in
the plots and they are compensated by the effects of multiple fields
and processes in the medium. That leads to the constraints on the
nucleon couplings even in the region undetectable in the case of HB
stars. Let us discuss several qualitative behaviors in order. Firstly,
we find that in the small $y_p$ region, the bounds are much weaker for
the hydrogen processes such as the H-H bremsstrahlung and the H$\gamma$
Compton. The other processes exhibit approximately symmetric bounds
with respect to $y_p$ and $y_n$. Secondly, we consider which processes
are dominant. This is a complex problem and involves several
independent effects. We find from the figure that the H-He
bremsstrahlung is dominant for $|y_p|<|y_n|$, whereas 
the $e$-H, $e$-He, and H-He bremsstrahlung contributions become
comparable for $|y_p|>|y_n|$. The reason for this is not immediately 
apparent. While the $e$-H bremsstrahlung depends on $y_p$ in the
vacuum, the $y_n$ dependence only appears through the multi-field
effects in the medium. That leads to somewhat more 
stringent constraints being placed on $y_p$ than $y_n$ by the $e$-H
bremsstrahlung process. As for the $e$-He bremsstrahlung, 
both $y_p$ and $y_n$ should be similarly constrained, but the actual
bound is weaker for $y_n$. This is because the cancellation occurs in
the $y_n$-dependent real part of the multi-field 
factor $c_\phi^{e\text{He}}$ near the resonance. Furthermore, as
mentioned in Section~\ref{sec:zphicphi}, the multi-field 
effect $c_\phi^X$ on the nucleus $X$ processes cancels the resonance
behavior coming from $z_\phi$, if $y_X\gtrsim y_e$ is satisfied, as in
the current hadrophilic scenario. This resonance cancellation also holds 
for $c_\phi^\text{HHe}$ associated with the H-He bremsstrahlung, 
since $c_\phi^\text{HHe}$ is obtained from the combination of
$c_\phi^\text{H}$ and $c_\phi^\text{He}$. After all, in the case of
the hadrophilic scalar, the contribution of H-He bremsstrahlung is
typically larger than those of other processes, but becomes less
significant at the resonance than that of the $e$-H
bremsstrahlung. As a result, comparable constraints on the Yukawa
couplings can be obtained from three bremsstrahlung processes. These
behaviors, which are the consequence of various medium effects, can
be seen in Figure~\ref{fig:ypyn}.

The inclusion of multi-field/coupling effects often leads to the
cancellation in production amplitudes and generates sharp dents and
slits for the parameter constraints. That was observed for the
lepton/hadron couplings in the previous section and can also be seen in
Figure~\ref{fig:ypyn}. For the HB stars, there is the slit 
around $y_p+y_n\simeq 0$, as already mentioned above. On the other
hand, the Sun presents a more complex situation, given its composition
of multiple nuclei and the inclusion of hydrogen. For $y_py_n>0$, the
bound from the H-He bremsstrahlung shows a dent approximately 
at $y_p\simeq y_n$ where the coefficient of the H-He amplitude
vanishes. The other processes or multi-field effects serve to
compensate this slit behavior. For $y_py_n<0$, we find two distinct
types of sharp dents. One corresponds to the helium-like processes 
observed in $L^{\text{HeHe}}$ and $L^{\text{He}\gamma}$. This is
similar to the HB star case in that a sharp dent appears in the direction 
where the helium Yukawa coupling vanishes. The other type of dent is
observed in the electron-like processes $L^{e\text{H}}$ and $L^{ee}$
in the hadrophilic model. These processes occur through the 
multi-field effect in the Sun, and the contributions from hydrogen
and helium nuclei cancel out 
when $y_\text{H}q_\text{H}n_\text{H}/m_\text{H}+ 
y_\text{He}q_\text{He}n_\text{He}/m_\text{He}\simeq0$ (see 
Eq.~\eqref{eq:cphieSun}). Under the condition of electrical neutrality, 
that corresponds to the relation between the nucleon couplings 
$y_n \simeq (n_\text{H}+n_e)/(n_\text{H}-n_e)y_p$ (the black dashed
line in Figure~\ref{fig:ypyn}).

\bigskip

\section{Summary}

In this paper, we have evaluated the scalar production in the
medium in the presence of multiple types of fields, particularly
including the contributions from heavy fields that are usually
ignored. By properly taking the plasmon mixing into account, we have
studied how and when multi-field effects are significant. For example,
while it has been previously stated that scalar processes are screened
in the low-energy regime, our finding indicated that this assertion is
valid only for the specific case of a single field/coupling. In the
presence of multiple fields and processes, the screening does not
occur or is weakened, depending on the type of processes and
model parameters. Such an observation means that the
inclusion of multiple (heavy) fields/couplings generally induces
non-negligible effects in the physics of scalar production in the medium.
Our evaluation was based on the fact that in the non-relativistic
regime, the Yukawa coupling of a scalar has the same form as the gauge
coupling to the photon. Since dark photons have the same form of
coupling as photons, the discussion can be parallel, and some
applications have already been discussed in the literature. On the
other hand, in the case of pseudo scalars such as the axion, the
coupling structure is different even in the non-relativistic regime,
and further study is needed to see whether the cancellation-like
behavior occurs.

As an application, we have analyzed the energy loss due to the scalar
emission in stellar interiors. The excessive energy emission affects the
evolution of stars and imposes some constraints on the property of the
scalar particle, such as the mass and couplings. The above medium
effects have been included into the evaluation of scalar emission from
various processes in the Sun and HB stars. In the stellar medium, 
where multiple nuclei such as hydrogen and helium exist
in addition to the electrons, the multi-field effects appear in various
ways, leading to distinct emission rates and resultant parameter
constraints. The weakened screening due to multi-field effects,
namely, the increased production rates at low energy, may not have
been a sizable contribution in view of the energy loss, but it could
be a crucial element from the perspective of particle number generation. 

The presence of heavier nuclei in the system, such as those observed
in white dwarfs, might significantly alter the previous analysis. However,
it would be necessary to incorporate the treatment of degenerate
particles into the current formalism. Moreover, it should be noted
that further modifications and corrections may be possible beyond the
scope of the present analysis. For example, the contribution from
the direct scalar coupling to two photons, higher-order temperature
corrections (arising from the velocity expansion of electrons), and
corrections due to the mass difference between the proton and neutron
could potentially affect the medium effects, especially in the
presence of cancellation among the leading-order contributions. It was 
also theoretically assumed in this paper that the scalar-plasmon
two-point function has a certain property, namely 
that $\Pi_{L\phi}(K)=\Pi_{\phi L}(K)$. This implies that all
contributing processes have the same phase, which may change in the
presence of CP-violating processes and couplings. Further
investigation into these corrections and assumptions is left for
future work.

\bigskip

\subsection*{Acknowledgments}
The work of Y.Y.\ was supported by the National Science and Technology
Council, the Ministry of Education (Higher Education Sprout Project
NTU-112L104022), the National Center for Theoretical Sciences of
Taiwan, and the visitor program of Yukawa Institute for Theoretical
Physics, Kyoto University. K.Y.\ is supported by the JSPS Grant-in-Aid
for Scientific Research KAKENHI Grant Number JP20K03949.

\bigskip\bigskip

\appendix

\section{Scalar self-energy with mixing}

We derive the general expression for the self-energy of a scalar 
field $\phi$ in the presence of mixing with other particles. We
consider the two-point Green function of $\phi$. When the sum of all
one-particle irreducible loop contributions is denoted 
as $\Pi_{\phi\phi}$, the one-particle irreducible two-point vertex 
function is obtained by inserting it, 
\begin{align}
  -\Gamma_{\phi\phi}^{(0)}{}^{-1} \;=\; 
  \Delta_\phi+\Delta_\phi\Pi_{\phi\phi}\Delta_\phi+
  \Delta_\phi\Pi_{\phi\phi}\Delta_\phi\Pi_{\phi\phi}\Delta_\phi + \cdots 
  \;=\; (\Delta_\phi^{-1}-\Pi_{\phi\phi})^{-1} ,
\end{align}
where $\Delta_\phi$ is the tree-level propagator of $\phi$. The
superscript $(0)$ means the level at which $\phi$ mixes with other
particles. If $\phi$ mixes with the particle $A$ (which, in the text,
is identified with the photon in the medium), then the mixing induces
additional contributions. When the sum of all one-particle irreducible
contributions to the $\phi$-$A$ two-point function is denoted 
by $\Pi_{\phi A}$, the above vertex function is additionally modified
by the contribution with two mixing parts,
\begin{align}
  -\Gamma_{\phi\phi}^{(2)}{}^{-1} &\;=\; 
  \Gamma_{\phi\phi}^{(0)}{}^{-1}\Pi_{\phi A}\Delta_A
  \Pi_{A \phi}\Gamma_{\phi\phi}^{(0)}{}^{-1}
  + \Gamma_{\phi\phi}^{(0)}{}^{-1}\Pi_{\phi A}\Delta_A
  \Pi_{AA}\Delta_A \Pi_{A \phi}\Gamma_{\phi\phi}^{(0)}{}^{-1}  
  + \cdots ,  \nonumber \\
  &\;=\; \Gamma_{\phi\phi}^{(0)}{}^{-1} 
  \Pi_{\phi A}(\Delta_A^{-1}-\Pi_{AA})^{-1}
  \Pi_{A \phi}\Gamma_{\phi\phi}^{(0)}{}^{-1} , \nonumber \\
  &\;\equiv\; \Gamma_{\phi\phi}^{(0)}{}^{-1} \Pi_\text{mix} 
  \Gamma_{\phi\phi}^{(0)}{}^{-1} . 
\end{align}
Here, $\Delta_A$ represents the tree-level propagator of $A$. We have
defined $\Pi_{AA}$ as the sum of all one-particle irreducible loop
contributions to the two-point function of $A$. While both external lines 
are $\Delta_\phi$ at the tree level, they are replaced 
by $-\Gamma_{\phi\phi}^{(0)}{}^{-1}$, as described above. Similarly, by
inserting $2n$ mixing, we have $-\Gamma_{\phi\phi}^{(2n)}{}^{-1}
=-\Gamma_{\phi\phi}^{(0)}{}^{-1} (-\Pi_\text{mix} 
\Gamma_{\phi\phi}^{(0)}{}^{-1})^n$. Summing up all the above
contributions, we obtain the two-point vertex 
function $\Gamma_{\phi\phi}$ with the mixing effect,
\begin{align}
  \Gamma_{\phi\phi} &\;=\; -\Big[
  \sum_{n=0} -\Gamma_{\phi\phi}^{(0)}{}^{-1} 
  (-\Pi_\text{mix}\Gamma_{\phi\phi}^{(0)}{}^{-1})^n \Big]^{-1}
  \;=\; \Gamma_{\phi\phi}^{(0)}+\Pi_\text{mix} \,.
  \label{eq:fullvertexfnc}
\end{align}
The self-energy of $\phi$ is defined by subtracting the tree-level
contribution from the two-point vertex 
function \eqref{eq:fullvertexfnc}. We then find the expression of the
full self-energy $\Pi_\phi$ including the mixing effect,
\begin{align}
  \Pi_\phi \;=\; \Pi_{\phi\phi} + \Pi_\text{mix}  \;=\; 
  \Pi_{\phi\phi} - \frac{\Pi_{\phi A}\Pi_{A\phi}}{\Pi_{AA}-\Delta_A^{-1}} \,.
\end{align}
The evaluation of these one-particle irreducible two-point functions
allows for the determination of the damping (thus
production/absorption) rate of the field $\phi$ in the medium.

\bigskip

\section{Production rates with multiple species of nuclei}

This appendix presents the explicit scalar production rates in the
case that there exist multiple species of nuclei in the
medium. Specifically, we assume three types of fields as in the Sun:
electrons ($e$), hydrogen (H), and helium (He). The following
processes are appropriate for 
the production of the scalar field $\phi$ at a temperature typical to
the solar interior ($X=$ H, He): 
(i) $e$-$e$ bremsstrahlung : $e+e\to e+e+\phi$, \ (ii) $X$-$X$
bremsstrahlung : $X+X\to X+X+\phi$, \ (iii) $e$-$X$ bremsstrahlung :
$e+X\to e+X+\phi$, \ (iv) $e$ Compton-like : $e+\gamma\to e+\phi$, \ 
(v) $X$ Compton-like~: $X+\gamma\to X+\phi$, and (vi) $X$-$X'$
bremsstrahlung ($X\neq X'$) : $X+X'\to X+X'+\phi$. 
The last process is available only when there are two or more 
nuclei including hydrogen in the medium. When $X, X' \neq$ hydrogen, the
non-relativistic amplitude of $X$-$X'$ bremsstrahlung in the vacuum
without mixing is zero as discussed in the text.

Extending the results of Section 2 to the case of two species of heavy
particles, the $\phi$ production rate for each process is evaluated 
including the medium effect factors $z_\phi$ and $c_\phi$:
\begin{align}
  \Gamma_\text{prod}^{ee} &\,=\,
  z_\phi c_\phi^e
  \frac{q_e^4 y_e^2 n_e^2 T^{1/2}k^2J(\omega)}{60\pi^{3/2} m_e^{5/2}\omega^5}\,,
  \\
  \Gamma_\text{prod}^\text{XX} &\,=\,
  z_\phi c_\phi^X  \frac{q_X^4 y_X^2 n_X^2 
T^{1/2}k^2J(\omega)}{60\pi^{3/2} m_X^{5/2}\omega^5} \,, \\
  \Gamma_\text{prod}^{eX} &\,=\, 
  z_\phi c_\phi^{eX}
 \frac{q_X^2q_e^2 n_e n_X m_e^{1/2}}{3(2\pi)^{3/2}}
  \Big(\frac{y_e}{m_e}  -\frac{y_X}{m_X}\Big)^2
 \frac{k^2 I(\omega)}{T^{1/2}\omega^5}  \,,
  \label{eq:eXBrems}   \\
  \Gamma_\text{prod}^{e\gamma} &\,=\, 
  z_\phi c_\phi^e \frac{q_e^2 y_e^2 n_e f_\phi k^3}{6\pi m_e^2 \omega^3} \,, \\
  \Gamma_\text{prod}^{X\gamma} &\,=\, 
  z_\phi c_\phi^X \frac{q_X^2 y_X^2 n_X f_\phi k^3}{6\pi m_X^2 \omega^3} \,, \\
  \Gamma_\text{prod}^{XX'} &\,=\, 
  z_\phi c_\phi^{XX'}
  \frac{q_X^2q_{X'}^2 n_X n_{X'} m_X^{1/2}m_{X'}^{1/2}}{3(2\pi)^{3/2}
(m_X+m_{X'})^{1/2}}
  \Big(\frac{y_X}{m_X}  -\frac{y_{X'}}{m_{X'}}\Big)^2
 \frac{k^2 I(\omega)}{T^{1/2}\omega^5} \,.
  \label{eq:XXpBrems}
\end{align}
\begin{align}
  c_\phi^e &\,=\, \Big|\,1+\frac{1}{\omega^2}\!\sum_{a=\text{H},\,\text{He}}\!
  \frac{q_ey_a - q_ay_e}{y_em_a} \Big[ 
  q_an_a  -i \frac{q_a}{m_a} P_a  +i \Big(\frac{q_e}{m_e}
  -\frac{q_a}{m_a}\Big)P_{ea} \Big]
  \nonumber \\
  & \hspace*{2cm} 
  -\frac{i}{\omega^2} \Big[\,\frac{q_e}{y_e}
\Big(\frac{y_\text{H}}{m_\text{H}}-\frac{y_\text{He}}{m_\text{He}}\Big) 
-\Big(\frac{q_\text{H}}{m_\text{H}}-\frac{q_\text{He}}{m_\text{He}}\Big) \Big] 
\Big(\frac{q_\text{H}}{m_\text{H}}-\frac{q_\text{He}}{m_\text{He}}\Big)P_\text{HHe}
  \,\Big|^2  ,
  \label{eq:cphieSun}
  \\[1mm]
  c_\phi^\text{H} &\,=\, \Big|\,1+\frac{1}{\omega^2}\!\sum_{a=e,\,\text{He}}\!
  \frac{q_\text{H}y_a-q_ay_\text{H}}{y_\text{H}m_a}
\Big[ q_a n_a -i \frac{q_a}{m_a} P_a  
+i\Big(\frac{q_\text{H}}{m_\text{H}}-\frac{q_a}{m_a}\Big)P_{a\text{H}} \Big] 
  \nonumber \\
  & \hspace*{2cm} 
  -\frac{i}{\omega^2} \Big[\,\frac{q_\text{H}}{y_\text{H}}
\Big(\frac{y_e}{m_e}-\frac{y_\text{He}}{m_\text{He}}\Big) 
-\Big(\frac{q_e}{m_e}-\frac{q_\text{He}}{m_\text{He}}\Big) \Big] 
\Big(\frac{q_e}{m_e}-\frac{q_\text{He}}{m_\text{He}}\Big)P_{e\text{He}}
  \,\Big|^2  ,
 \\[1mm]
  c_\phi^\text{He} &\,=\, \Big|\,1+\frac{1}{\omega^2}\!\sum_{a=e,\,\text{H}}\!
  \frac{q_\text{He}y_a-q_ay_\text{He}}{y_\text{He}m_a}
\Big[ q_a n_a  -i \frac{q_a}{m_a} P_a  
+i\Big(\frac{q_\text{He}}{m_\text{He}}-\frac{q_a}{m_a}\Big)P_{a\text{He}} \Big]
  \nonumber \\
  & \hspace*{2cm} 
  -\frac{i}{\omega^2} \Big[\,\frac{q_\text{He}}{y_\text{He}}
\Big(\frac{y_e}{m_e}-\frac{y_\text{H}}{m_\text{H}}\Big) 
-\Big(\frac{q_e}{m_e}-\frac{q_\text{H}}{m_\text{H}}\Big) \Big] 
\Big(\frac{q_e}{m_e}-\frac{q_\text{H}}{m_\text{H}}\Big)P_{e\text{H}} \,\Big|^2 ,
  \\[1mm]
  c_\phi^{e\text{H}} &=\, \Big|\,  1+\frac{1}{\omega^2}
  \frac{q_ey_\text{H}-q_\text{H}y_e}{y_e m_\text{H}-y_\text{H} m_e}
  \!\sum_{a=e,\text{H},\text{He}}\!\!
  \Big(q_an_a - i\frac{q_a}{m_a}P_a\Big) 
  \nonumber \\
  & \hspace*{2cm}
  -\frac{B}{\omega^2 m_\text{He}(y_em_\text{H}-y_\text{H}m_e)} \Big[\,
  q_\text{He}n_\text{He}  -i \frac{q_\text{He}}{m_\text{He}}P_\text{He}
  \nonumber \\[1mm]
  & \hspace*{4cm}
  -i\Big(\frac{q_e}{m_e}-\frac{q_\text{He}}{m_\text{He}}\Big)P_{e\text{He}}
  -i\Big(\frac{q_\text{H}}{m_\text{H}}
  -\frac{q_\text{He}}{m_\text{He}}\Big)P_\text{HHe} \,\Big] \,\Big|^2 ,
  \\[1mm]
  c_\phi^{e\text{He}} &=\, \Big|\,  1+\frac{1}{\omega^2}
  \frac{q_ey_\text{He}-q_\text{He}y_e}{y_e m_\text{He}-y_\text{He} m_e}
  \!\sum_{a=e,\text{H},\text{He}}\!\!
  \Big(q_an_a - i\frac{q_a}{m_a}P_a\Big) 
  \nonumber \\
  & \hspace*{2cm}
  +\frac{B}{\omega^2 m_\text{H}(y_em_\text{He}-y_\text{He}m_e)} \Big[\,
  q_\text{H}n_\text{H}  -i \frac{q_\text{H}}{m_\text{H}}P_\text{H}
  \nonumber \\
  & \hspace*{4cm}
  -i\Big(\frac{q_e}{m_e}-\frac{q_\text{H}}{m_\text{H}}\Big)P_{e\text{H}}
  +i\Big(\frac{q_\text{H}}{m_\text{H}}
-\frac{q_\text{He}}{m_\text{He}}\Big)P_\text{HHe} \,\Big] \,\Big|^2 ,
  \\[1mm]
  c_\phi^{\text{HHe}} &=\, \Big|\,  1+\frac{1}{\omega^2}
  \frac{q_\text{H}y_\text{He}-q_\text{He}y_\text{H}}{y_\text{H} m_\text{He}
-y_\text{He} m_\text{H}}
  \!\sum_{a=e,\text{H},\text{He}}\!\!
  \Big(q_an_a - i\frac{q_a}{m_a}P_a\Big) 
  \nonumber \\
  & \hspace*{2cm}
  -\frac{B}{\omega^2 m_e(y_\text{H} m_\text{He}-y_\text{He}m_\text{H})}
  \Big[\,  q_e n_e  -i \frac{q_e}{m_e}P_e
  \nonumber \\[1mm]
  & \hspace*{4cm}
  +i\Big(\frac{q_e}{m_e}-\frac{q_\text{H}}{m_\text{H}}\Big)P_{e\text{H}}
  +i\Big(\frac{q_e}{m_e}
  -\frac{q_\text{He}}{m_\text{He}}\Big)P_{e\text{He}} \,\Big]
    \,\Big|^2 .
\end{align}
\begin{align}
  B \,=\, m_e(q_\text{H}y_\text{He} -q_\text{He}y_\text{H}) 
  + m_\text{H}(q_\text{He}y_e -q_ey_\text{He}) 
  + m_\text{He}(q_ey_\text{H} -q_\text{H}y_e) \,.
\end{align}
The functions $P_i$, $P_{ij}$, $I(\omega)$, and $J(\omega)$ are defined 
in \eqref{eq:Pi}--\eqref{eq:fncJ}. As for the medium effect 
factors, $z_\phi$ is common to all processes, but the multi-field
factors $c_\phi$ are modified by the existence of two types of nuclei
compared to \eqref{eq:cphie}--\eqref{eq:cphieX}. In the above
expression of $c_\phi^{e\text{H}}$, $c_\phi^{e\text{He}}$, and 
$c_\phi^{\text{HHe}}$ with the bremsstrahlung, the 
term $\sum q_an_a$ vanishes under the condition of electrical
neutrality in the medium. The $\phi$ production rates in the Sun can
be obtained by substituting specific values for the electron, hydrogen,
and helium in the above general formulae: 
$q_e=-e$, $\,q_\text{H}=e$, $\,q_\text{He}=
2e$, $\,m_\text{H}=m_N$, $\,m_\text{He}=4m_N$ ($m_p=m_n\equiv m_N$), and
the Yukawa couplings $y_\text{H}=y_p$, $\,y_\text{He}=2y_p+2y_n$.

\clearpage

\newcommand{\arxivfont}{\rmfamily}
\bibliographystyle{utphys}
\bibliography{references}

\end{document}